\documentclass[12pt]{article}
\pdfoutput=1
\usepackage{booktabs} 
\usepackage{array} 
\usepackage[table,xcdraw]{xcolor} 
\usepackage{amsmath,amssymb}
\usepackage{graphicx}
\usepackage{fancyhdr}
\usepackage{bm, color}
\usepackage{slashed,amsthm,amsfonts,empheq}
\usepackage[caption=false]{subfig}
\usepackage{hyperref}
\usepackage{booktabs}
\usepackage{multirow}
\usepackage[left=2cm,right=2cm,top=2cm,bottom=2cm,a4paper]{geometry}
\usepackage{amsmath,amssymb}
\usepackage{graphicx}
\usepackage{cite,fancyhdr}
\usepackage{bm, color}
\usepackage{amssymb,amsfonts,slashed,amsthm,amsmath,graphicx,soul,empheq}
\usepackage[caption=false]{subfig}
\usepackage{hyperref}
\newcommand{\Slash}[1]{{\ooalign{\hfil#1\hfil\crcr\raise.167ex\hbox{/}}}}

\newcommand{\beq}{\begin{equation}}  \newcommand{\eeq}{\end{equation}}
\newcommand{\bef}{\begin{figure}}  \newcommand{\eef}{\end{figure}}
\newcommand{\bec}{\begin{center}}  \newcommand{\eec}{\end{center}}
  
\newcommand{\laq}[1]{\label{eq:#1}}  

\newcommand{\Eq}[1]{Eq.(\ref{eq:#1})}

\def\({\left(}
\def\){\right)}

\def\O{\mathcal{O}}
\def\U{\mathop{\rm U}}

\newcommand{\OR}{~{\rm or}~}

\newcommand{\GEV}{ {\rm GeV} }

\def\d{\delta}

\def\f{\phi}

\def\s{\sigma}

\def\x{\xi}

\def\F{\Phi}

\def\*{\dagger}

\usepackage[affil-it]{authblk}
\usepackage[left=2cm,right=2cm,top=2cm,bottom=2cm,a4paper]{geometry}

\begin{document}
\begin{titlepage}
\begin{center}
%\allowdisplaybreaks
\setcounter{footnote}{0}
\setcounter{figure}{0}
\setcounter{table}{0}

\hspace{3cm}

\hspace{3cm}

\hspace{3cm}

\hspace{3cm}

{\Large\bf 
 Non-Scaling Topological Defects and Gravitational Waves in Higgs Portal}

\vskip .75in

{ \large    Wen Yin}

\vskip 0.25in

\begin{tabular}{ll}
& {\em Department of Physics, Tokyo Metropolitan University,} \\
&{
Minami-Osawa, Hachioji-shi, Tokyo 192-0397 Japan} \\[.3em]
&\\
& 
\end{tabular}

\begin{abstract}
One of the simplest extensions of the Standard Model is the Higgs portal extension involving a dark Higgs. Dark sectors that include dark matter candidates, WIMPs, axions, and dark photons, can naturally have this type of interaction, where the dark Higgs is charged under some symmetry, which may or may not be spontaneous broken by the vacuum expectation value.
In this paper, using lattice simulations, I show that if the reheating temperature of the Universe is sufficiently high, topological defects such as domain walls and cosmic strings associated with these symmetries are naturally formed even if the symmetries are never restored due to negative thermal mass squareds. This occurs due to the early Universe's non-adiabatic oscillation of the Higgs around the onset of oscillation, which overshoots the origin, and tachyonic instability that enhances fluctuations. 
The gravitational waves generated by these topological defects may be very significant due to the energetic processes induced by matter effects in the hot and dense Universe irrelevant to the typical energy scale of the dark sector in the vacuum or whether the symmetry is broken in the vacuum. 
Alongside earlier studies such as usual phase transition, melting domain walls and melting cosmic strings scenarios that assume a symmetric phase in the early Universe, 
the Higgs portal models naturally predict local overdensities from topological defects, which may induce miniclusters and primordial black holes, as well as the gravitational waves. 
These phenomena provide novel opportunities to search for such scenarios. 
I also perform various numerical simulations for the relevant topic including melting domain walls and cosmic strings with inflationary and Gaussian fluctuations, for comparison— which have not been performed previously.
\noindent
\end{abstract}

\end{center}
\end{titlepage}
\setcounter{footnote}{0}
\setcounter{page}{1}

%\maketitle

\section{Introduction}

Higgs portal extensions of the Standard Model (SM) provide one of the simplest frameworks to connect the SM Higgs sector with new physics such as dark matter or other hidden sectors~\cite{Silveira:1985rk,Burgess:2000yq}. A slight modification of the model has been studied in the WIMP dark matter~\cite{Barger:2008jx,Barger:2010yn,Gonderinger:2012rd,Ishiwata:2018sdi,Cline:2019okt,Grzadkowski:2020frj,Abe:2020iph,Abe:2021nih,Abe:2021jcz}, 
in the context of axion/dark photon dark matter production via phase transitions~\cite{Nakayama:2021avl}, the UV model of a CP-even ALP~\cite{Sakurai:2021ipp}, 
heavy dark matter production~\cite{Azatov:2021ifm,Azatov:2022tii}, electroweak baryogenesis~\cite{Barger:2008jx,Cho:2021itv,Blasi:2023rqi}, and collider physics~\cite{Chen:2019ebq,Grzadkowski:2020frj,Abe:2021nih,Bhattacherjee:2021rml,Sakurai:2021ipp,Sakurai:2022cki,Haghighat:2022qyh} etc. In these scenarios, a dark scalar (“dark Higgs”) couples through the Higgs portal operator, allowing for a rich cosmological evolution in the early Universe. If the dark scalar spontaneously breaks a discrete or continuous symmetry, topological defects such as domain walls (DWs) or cosmic strings (CSs) can naturally arise. These defects are known to leave potentially observable imprints on the Universe, including the production of gravitational waves (GWs).

Traditionally, it is assumed that symmetry restoration at high temperatures in the early Universe is necessary for topological defects to form later, once the symmetry is spontaneously broken at lower temperatures. This picture is often referred to as the Kibble mechanism~\cite{Kibble:1976sj,Vilenkin:1984ib,Vilenkin:2000jqa}, in which large regions of space independently choose symmetry-breaking vacua, giving rise to defects. However, more recent studies have suggested alternative pathways for defect formation that do not rely on a fully restored symmetric phase. For instance, \cite{Gonzalez:2022mcx,Kitajima:2023kzu} pointed out that if strong scale-invariant fluctuations, which are the fluctuations generated during inflation, are present, those fluctuations alone can effectively cover regions of field space near the origin, mimicking the usual scenario of symmetry restoration in the context of DWs and CSs. This is in contrast to the conventional understanding, where population bias would make the topological defects unstable. The key observation is the superhorizon correlation: some domains are generated at superhorizon sizes, similar to the voids in density perturbations, and so they are long-lived.

If the dark Higgs field has a negative portal coupling, it may develop a large potential minimum in the early Universe when the temperature is high. This phenomenon is used to propose melting CSs~\cite{Emond:2021vts} and melting DW scenarios~\cite{Babichev:2021uvl}, showing the absence of the DW problem due to decreasing tension over time (see also some early studies on topological defects with time-dependent tension~\cite{Yamaguchi:2005gp,Ichikawa:2006rw}). It was analytically shown that the GW signal~\cite{Emond:2021vts,Babichev:2021uvl,Babichev:2023pbf}, and the DW one is studied numerically by performing lattice simulations, noting the simulation-friendly setup of time-varying tension~\cite{Dankovsky:2024ipq} with the GW spectra shown, taking into account Minkovski fluctuations and thermal fluctuations at the zero mass limit. The setups so far assumed that initially the dark Higgs field stays at the origin, namely the system is initially in the symmetric phase. 

{More recently, a very weakly coupled model has been proposed as a natural UV model of the QCD axion, which takes account of the impact of the negative Higgs portal for the dark matter production~\cite{Yin:2024txg}. This model assumes that the wave function renormalization constant of the Peccei-Quinn Higgs is very large, instead of assuming a large Higgs mass to explain the high Peccei-Quinn scale, while keeping other parameters at $\mathcal{O}(1)$ in units of the electroweak scale. This approach can alleviate the fine-tuning associated with the electroweak scale and address the quality problem of the Peccei-Quinn symmetry. When the thermally induced Higgs Compton wavelength becomes comparable to the Hubble parameter, the finite-temperature induced minimum attains a field value around the Planck scale in a large parameter region. Thus, if topological defects are formed during this era, very strong gravitational waves (GWs) may be emitted. This UV model is applicable to generic dark Higgs models, i.e., taking the wave function renormalization constant to be very large to achieve a large vacuum expectation value, while other parameters remain at $\mathcal{O}(1)$ in units of the typical energy scale of the model. However, reaching the symmetric phase in the early Universe is difficult due to the feeble interactions. 
}

In this work, I uncover yet another scenario in which topological defect formation can occur even without the averaged field spending long (or any) time in a symmetric phase due to the negative thermal mass squared. Specifically, assuming a high enough reheating temperature,  I show that the combination of a negative thermal mass squared and oscillations of the dark Higgs field inevitably draws the field back to the origin at least once due to effect of the cosmic expansion. In passing through the origin, a transient tachyonic instability can dramatically enhance field fluctuations, seeding the formation of topological defects such as DWs or CSs. Crucially, this implies that even if the high-temperature state does not restore the symmetry in the usual sense, topological defects can still materialize. 
Those  subhorizon sized topological defects are interestinly long-lived similar to the melting DWs or CSs. 

Importantly, these topological defects form during a period when the Universe is hot and dense, resulting in defect tensions that are significantly larger than their zero-temperature values. This means that defects can exist even when symmetry is restored at zero temperature. In the case of domain walls (DWs) or cosmic strings (CSs), the tension—defined as the mass per unit area or length—is dominated by thermal effects and scales with the temperature at the time of formation. In some models, such as \cite{Yin:2024txg}, the thermally induced tension can be extremely large in the early Universe. The subsequent dynamical evolution of these defects can produce gravitational waves (GWs) with characteristic frequencies and amplitudes determined by the formation time and tension scale, as will be discussed below.
 
I also perform lattice simulations of the relevant scenarios, such as melting DWs with Gaussian and scale invariant fluctuations and melting cosmic strings, which have not been performed yet, for comparison of the dynamics and GWs in the appendix.

\section{Setup}

\paragraph{Potential of the system in physical notation}
\beq \label{pot} 
V \supset \lambda_P |\F|^2 |h|^2 - \mu_h^2 |h|^2 - m^2_\F |\F|^2 + \lambda_h |h|^4 + \lambda |\F|^4, 
\eeq
where $h$ is the SM-like Higgs doublet field. $\F$ is a dark Higgs field, which can be either a complex scalar or a real scalar, depending on whether the symmetry it breaks is global $U(1)$ or $Z_2$. In other words, we take 
\beq 
|\F|^2 = \sum_{i}^n \frac{\f_i^2}{2}, 
\eeq 
with $n=1$ or $2$. 
We consider a cosmological history in which the $H$ mass is always smaller than the cosmic temperature. 
Under this assumption, the thermal mass contribution to $\F$ is given by
\beq 
m^2_{\rm th} \sim \frac{\lambda_P}{6} T^2. 
\eeq 
See \cite{Yin:2024txg} for further discussion and derivation, allowing $\lambda_P < 0$ and a more precise form.  
Importantly, we assume 
\beq
\lambda_P < 0,
\eeq
In the thermal environment, by neglecting the zero-temperature mass, the effective potential becomes
\beq
V_{\rm eff} \simeq m_{\rm th}^2 |\F|^2 + \lambda |\F|^4. 
\eeq

\begin{figure}[!t] 
    \begin{center}
        \includegraphics[width=150mm]{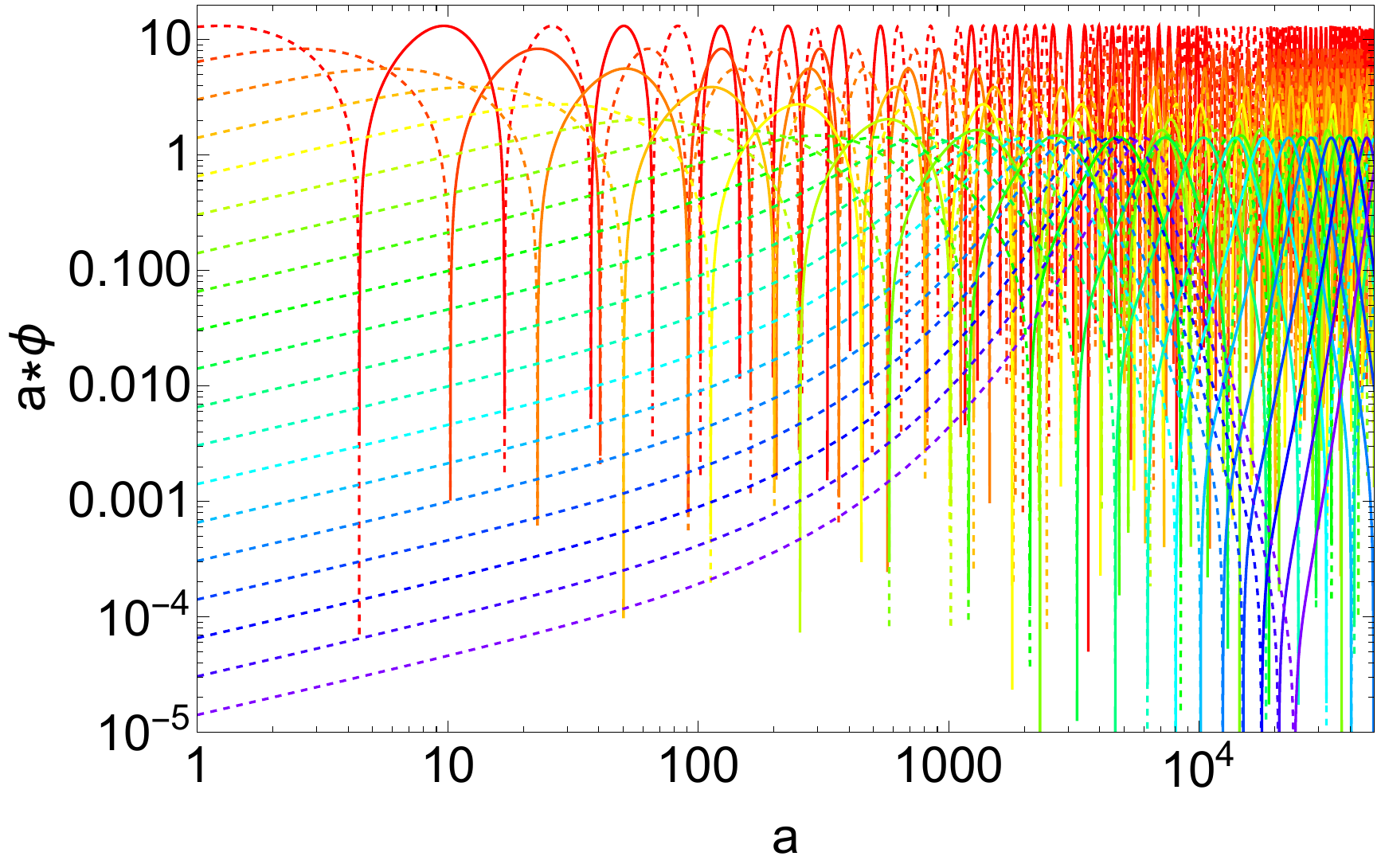}
    \end{center} 
    \caption{Solution of the field equation {\it without fluctuations}. Here, $\phi$ multiplied by the scale factor $a$ is plotted against varying scale factor which is normalized to be unity at the beginning. Several initial conditions in the radiation-dominated era with negative thermal mass squared are used. The dashed lines denote $a |\phi|$ when $\phi$ is positive, while solid lines denote the same quantity when $\phi$ is negative. See the main text for the setup of the calculation.
    }
    \label{fig:1} 
\end{figure}

If the temperature is high enough, the potential minimum becomes temperature-dependent:
\beq \label{finite}
|\F_{\rm min}|_T^2 = \frac{|\lambda_P|}{24\lambda} T^2.
\eeq 
Given that 
\beq 
T \propto a^{-1},
\eeq 
which is also the assumption throughout the paper, the thermal mass scales with $a^{-1}$, where $a$ is the scale factor. In the radiation-dominated or matter-dominated Universe, $\lim_{T\to \infty} |m_{\rm th}|/H \to 0$ since $H \propto a^{-2} \OR a^{-3/2}$. Therefore, if the temperature is high enough, Hubble friction is more important than the mass.  
Since the thermal mass in this setup, and in generic thermal field theory, is typically larger than the thermalization rate, the initial condition that friction is more important than the thermal mass implies that the dark Higgs sector is not thermalized at the relevant period.\footnote{It is also possible to have different setups of the fluctuations by introducing new particles/higher-dimensional terms or by considering incomplete thermal production (or freeze-in)~\cite{Nakayama:2021avl}. In the former case, we may have a complete thermal distribution, while in the latter case, the distribution depends on the reaction but usually has an amplitude smaller than the thermal one. Although they will provide additional possibilites for the initial conditions in the simulation, I do not consider them in this paper (see c.f. Ref.\,\cite{Dankovsky:2024ipq}).} 

Since I consider the early Universe dynamics when the thermal mass $m_{\rm th}^2$ is very large, I will neglect the vacuum mass $m_{\F}^2$ in the following.
Indeed, what I will show does not depend on the sign of $m^2_\F$, and thus, this dark Higgs may or may not break the symmetry in the vacuum. 

\begin{figure}[!t] 
    \begin{center}
        \includegraphics[width=150mm]{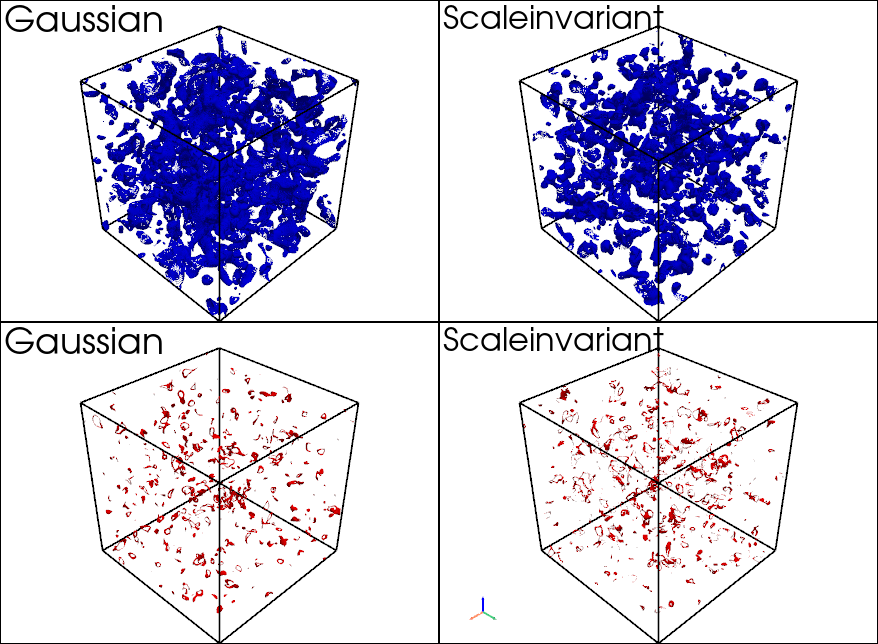}
    \end{center} 
    \caption{Snapshot of the lattice simulation at $\tau=101d^{-1}$. The top panels (blue contours) represent the case for the $Z_2$ symmetric potential, while the lower panels represent the $U(1)$ case. The left panels have Gaussian initial fluctuations, while the right panels have scale-invariant fluctuations. Here, $\bar{\F} \neq 0$ as the initial fluctuation.}
    \label{fig:2} 
\end{figure}

\begin{figure}[!t] 
    \begin{center}
        \includegraphics[width=150mm]{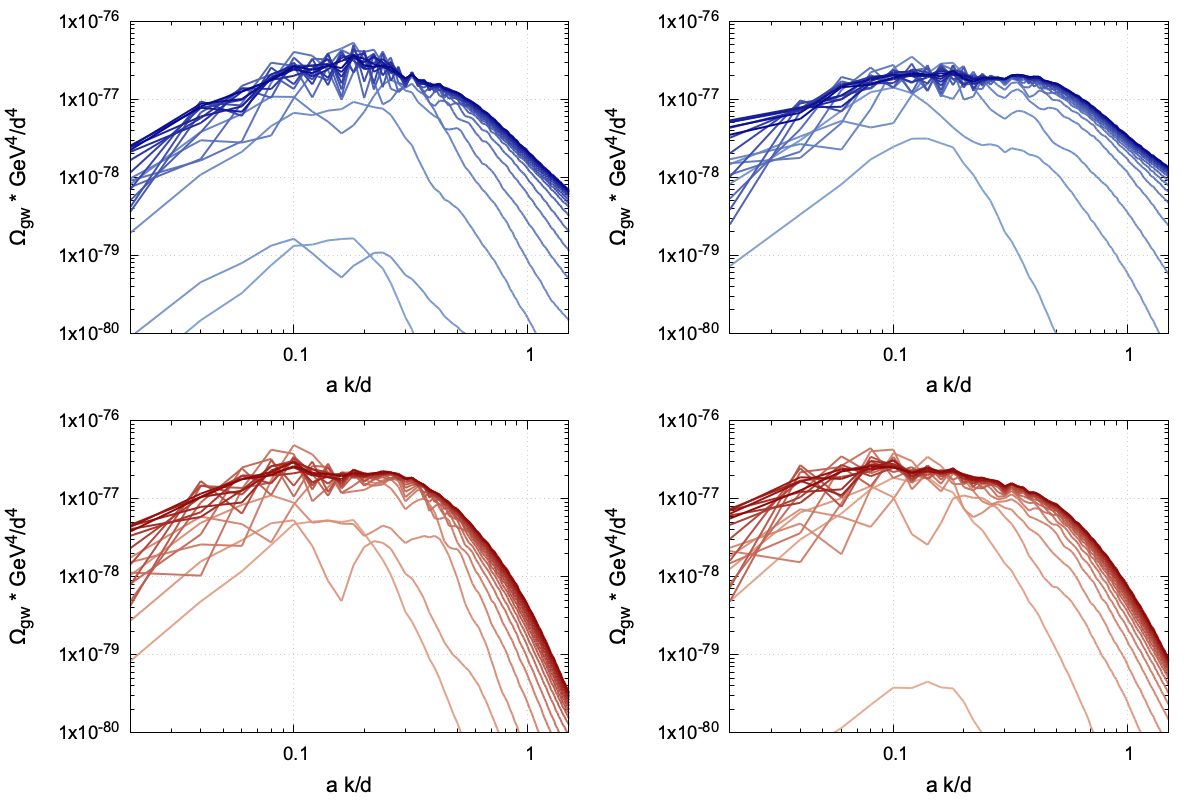}
    \end{center}
    \caption{Time evolution of the GW spectra. Each panel corresponds to those in Fig.~\ref{fig:2}. The time evolution is taken from $\tau=1d^{-1}$ to $\tau=231d^{-1}$, with the plots shown from lighter to darker for each panel.}
    \label{fig:3} 
\end{figure}

\paragraph{Initial condition and simulations}
The initial conditions for fields can be decomposed as 
\beq
\f_i^0 = \bar{\f}^0_{i} + \delta \f^0_i,
\eeq
where $\bar{\f}^0_{i}$ is the homogeneous mode, and the fluctuation $\d \f _i$ satisfies
\beq
\langle \delta \f_i^0 \delta \f_j^0 \rangle = \delta_{ij} \int d \log k \, {\cal P}_{\delta \f}(k).
\eeq
For the reduced power spectrum, we consider two possibilities following \cite{Gonzalez:2022mcx, Kitajima:2023kzu}:
\beq
{\cal P}_{\delta \f}(k) = \frac{c k^3}{4\pi^2} \quad \text{(Gaussian)}, \quad \frac{H_{\rm inf}^2}{4\pi^2} \quad \text{(Inflationary)},
\eeq
where $H_{\rm inf}$ is the inflationary Hubble parameter. $c$ can be considered a dimension$^{-1}$ parameter. 
Which possibility to consider depends on the early cosmology and particle models. 

For instance in the minimal setup, the dark Higgs field can acquire a sizable Hubble-induced mass during inflation via the non-minimal coupling:
\beq
{\cal L} \supset - \xi |\F|^2 R \to \delta V = 3 \xi H_{\rm inf}^2 |\F|^2,
\eeq
where $R$ is the Ricci scalar. 
Depending on the size and sign of $\xi$, one has different natural initial conditions for $\F$ around the beginning of the Universe:
\begin{table}[htbp]
    \centering
    \begin{tabular}{|c|c|c|}
        \hline
        Non-minimal coupling & $|\bar{\F}_{\rm ini}|$ & ${\cal P}_{\delta \f}$ \\ \hline
        $\xi \lesssim -1$ & $\sim \sqrt{H_{\rm inf}^2 / (\lambda |\xi|)}$ & Gaussian \\ \hline
        $|\xi| \lesssim 1$ & $(H_{\rm inf}^4 / \lambda)^{1/4}$ & Scale-invariant \\ \hline
        $\xi \gtrsim 1$ & $0$ & Gaussian \\ \hline
    \end{tabular}
    \label{tab:xf_dependence}
\end{table}

In the setup of $|\xi| \gtrsim 1$, I assume that preheating phenomena do not significantly alter the fluctuations. This is the case where, soon after inflation, the inflaton mass becomes much larger than the Hubble parameter, which is usually satisfied for low-scale inflation. In large-field inflation, the discussion depends on the size of $\xi$ and the inflation model, e.g.,~\cite{Li:2022ugn}. For concreteness, I assume that (p)reheating ends soon after inflation and thus the Hubble induced mass gets suppressed soon after inflation.  In the case $|\x| \lesssim 1$ I assumed that inflation lasts long enough to have stochastic distributions see e.g. \cite{Yin:2024txg}. 

At the boundaries of the regime, we can also have different types of initial conditions~\cite{Takahashi:2020tqv, Kitajima:2023kzu}.
In fact, other than $\xi \gtrsim 1$, one has the symmetry spontaneously broken during inflation in the observable Universe, which is the focus of this paper and is actually a natural setup. When reheating ends, the non-minimal coupling is no longer important.  
In the cases $|\xi| \lesssim 1$ and $\xi \lesssim -1$, the setup naturally has $\F$ not at the origin initially. 
This setup is the focus of the main part, while the other possibility, including the melting topological defect scenario~\cite{Babichev:2021uvl, Emond:2021vts, Babichev:2023pbf, Dankovsky:2024ipq} corresponding to the $\xi \gg 1$ case, will be simulated in the appendix, as some scenarios have not been studied numerically so far. 

To address late-time physics,  
we can solve the equation of motion for the homogeneous mode:
\beq
\ddot{\f}_i + 3H \dot{\f}_i = -V_{{\rm eff},\f_i},
\eeq
i.e., by neglecting the gradient contribution. 
To solve the equation, I take $m_{\rm th}^2 \ll H^2$ and set $\dot{\f}_i = 0$ at conformal time $\tau=1$ as the initial condition, which is generic due to the slow-roll attractor behavior. 
With this initial condition, which models the previously discussed early Universe picture, 
it is sufficient to use a single field or radial mode, $\f \equiv \frac{\sum_{i} \bar{\f}_{i}^0 \f_i}{|\bar{\F}^0|}$, to solve the equation. 
I find in Fig.~\ref{fig:1} that $\f$ always overshoots the origin for any initial $\bar\f^0$. This is also checked by assuming a matter-dominated background expansion. This occurs because if the adiabatic invariant is conserved, the oscillation amplitude scales the same as $|\F_{\rm min}|_T \propto a^{-1}$, but the initial adiabaticity is violated at the onset of oscillation. 
I cannot even find a fine-tuned initial condition to avoid the overshoot. 
This may be becasue tuning the initial position in the early Universe exactly following the minimum is not possible due to the time-evolution while the field is frozen by Hubble friction. 

Such overshoots around the potential origin imply the possibility of enhancing fluctuations non-linearly via non-perturbative effects such as tachyonic instability~\cite{Felder:2000hj,Felder:2001kt}. Additionally, dynamics passing through the symmetry-enhancing point imply that particle/topological defect production can occur. 
Even if we neglect the above effects, we have parametric resonance~\cite{Kofman:1994rk, Kofman:1997yn, Dufaux:2006ee, Amin:2018kkg}, since the amplitude very large in the unit of $|\F_{\rm min}|_T$. 
To investigate this, I modified {\tt CosmoLattice}~\cite{Figueroa:2020rrl,Figueroa:2021yhd} by incorporating the scale factor-dependent mass squared and the non-trivial initial conditions in the two/single scalar field system for the numerical simulation, including the fluctuations. 
While the setup and details are provided in the appendix, the numerical results are shown in Fig.~\ref{fig:2}. I assume in the figure that $\bar{\f}_i^0$ is around the origin, but $|\f_i^0 + \delta \f^0|$ is typically far from the origin.  
In the left panels, I show the case of $Z_2$ symmetry, and the right panels represent the $U(1)$ symmetry case. In the top and bottom panels, I denote the Gaussian case and scale-invariant case, respectively, to account for the possibilities of the initial condition. 
Interestingly, topological defect formations are clearly observed despite the system never reaching the symmetric phase. The size of the defects is much smaller than the Hubble volume. Indeed, each side of the box in Fig.~\ref{fig:2} contains roughly two Hubble horizons. 
They are small because their formation occurs in a limited field space when $\f_i$ overshoots the origin in the first few oscillations. 
Note that at the overshoot, the symmetry is not restored because $\dot{\phi}_i$ has a preferred direction.\footnote{I also note that by fine-tuning the fluctuations, a few overshoots may cause the field to stall at the hilltop, in which case symmetry restoration and the formation of large-sized topological defects are more likely, especially in the scale-invariant fluctuation case. I will not discuss this fine-tuned scenario.} 

Since the topological defects have tensions,
\beq 
\sigma \sim \frac{|m_{\rm th}^3|}{\lambda}, \quad \mu \sim \frac{|m_{\rm th}^2|}{\lambda}, \laq{tension}
\eeq
which scale with $a^{-D}$, where $D=3$ for DWs and $D=2$ for CSs, respectively, the loops of the topological defects are relatively long-lived compared to conventional ones.  
This can be understood from momentum conservation by coarse-graining the small loop as a ``particle," which has momentum $p \sim s \sigma v \sim s \s$ or $p \sim l \mu v \sim l\mu$, where $s$ is the surface area of the DW loop, $l$ is the length of the string loop, and $v$ is the velocity, assumed to be around the speed of light since we consider subhorizon loops. The momentum should redshift as $p \propto a^{-1}$ if other momentum losses are neglected, due to cosmic expansion. Therefore, we get $s \propto a^2$ and $l \propto a$, which implies that they are frozen to have a typical fraction in the comoving box. Therefore, they are long-lived compared to conventional defects. 

I consider this phenomenon to be generic and irrelevant to small fluctuations. In the case where we tune the fluctuations to be small, the Higgs tends to overshoot the origin more than once. Again, the Higgs fields setles into the potential minimum due to the fluctuation productions, and therefore the aforementioned topological defects form when the Higgs fields pass through the potential hilltop because the fluctuations should be large.  
I also observe that the enhancement of inflationary fluctuations due to tachyonic instability can make the inflationary stability against population bias~\cite{Gonzalez:2022mcx, Kitajima:2023kzu} more robust, as shown in the appendix.

\section{GW Production}

Since topological defects are formed, which are known to efficiently induce GWs, their presence is significant. In particular, independent of the vacuum mass $m_\F^2$, the typical energy scale for the tensions of the topological defects in \Eq{tension} is determined by finite-temperature effects and can be very large at the time of formation. Note that even if the Higgs portal coupling is very small, the tension can be large if the temperature is sufficiently high and if $\lambda$ is small. See \cite{Yin:2024txg} for an example.

This motivated me to study the GWs, which could be an important probe of the Higgs portal to the dark Higgs sector. The result is shown in Fig.\ref{fig:3}.

Interestingly, the peak frequency and the GW spectrum can be approximated by the usual ones from topological defects in the standard scaling solution. However, we use the tensions and Hubble parameters at the time of formation to estimate the frequency and the peak energy density parameter.

Thus, the peak comoving momentum is estimated as  
\beq
a k_{\rm peak} \sim \left.\,a H\right|_{T=T_{\rm osc}},
\eeq
in both cases because we only have a single typical energy scale, which is $m_{\rm th} \sim H$ at this moment. This agrees well with the simulation where the topological defects form at $\tau = \mathcal{O}(10)/d$, and therefore the peak frequency is around $\mathcal{O}(0.1) d$, which is consistent.

In addition, the peak energy density parameter, which is $\partial_{\log{k}} \rho_{\rm gw}/\rho_{\rm tot}|_{t=t_{\rm form}}$, is estimated in \cite{Hiramatsu:2013qaa} for DWs (see, e.g., the upper panels of Fig.~\ref{fig:3}) as
\beq
\Omega^{\rm DW}_{\rm gw,peak} \sim \left.\,\frac{\sigma^2}{24\pi M_{\rm pl}^4 H^2}\right|_{\tau=\tau_{\rm form}} \approx 10^{-76} \left(\frac{40}{\tau_{\rm form}}\right)^2 \frac{d^{4}}{\GEV^4},
\eeq
where I neglected various order-one factors. In the second approximation, I used the parameter for the numerical simulation, which is consistent with the numerical simulation, Fig.\ref{fig:3}.

For the CS case, I recast the result from~\cite{Emond:2021vts}:
\beq
\Omega^{\rm ST}_{\rm gw,peak} \sim \left(\frac{G \mu_{\rm t_{\rm form}}}{10^{-2}}\right)^2 \approx 10^{-75} \left(\frac{70}{\tau_{\rm form}}\right)^4 \frac{d^4}{\GEV^4},
\eeq
which also agrees well with the numerical one.

The $k$ and $\Omega_{\rm gw}$ are, of course, not the same as those for the present Universe, which is normalized at the initial time where I take $\tau=a/d=1/d$. To relate the numerical results to phenomenology for GW observations, we can use the typical scale instead of the machine unit $d$, and employ the redshift relation provided in, e.g., Ref.~\cite{Kitajima:2023cek}.

I also checked that with a large initial amplitude of $|\bar{\F}^0| \gg |m_{\rm th}|/\sqrt{\lambda}, |\delta \F|$ with thermally induced symmetry breaking or a constant mass term, the GWs are weaker within the simulation time. One could consider this as the symmetry restoration process. As long as we consider the thermal mass-dominated case, the oscillation amplitude of $\F$ via the quartic potential scales as $a^{-1}$, similar to $m_{\rm th}$, and thus the phase transition never occurs. However, later when the Universe cools down sufficiently in realistical setup, the vacuum mass becomes important. This drives the symmetry breaking. Then we have the usual phase transition and topological defect formation, as well as the GWs. Thus the phenomena observed in the main part is for $|\bar{\F}^0| < \mathcal{O}(m_{\rm th}/\sqrt{\lambda})$ at the onset of oscillation. 

\section{Conclusions and discussion}

I have shown that  if the reheating temperature of the Universe is sufficiently high, topological defects such as domain walls and cosmic strings associated with these symmetries are naturally formed even if the symmetry is never restored  due to a negative thermal mass squared or even restored in the vacuum. 
Thus, alongside other well-studied phenomena such as  conventional phase transition-induced defects, ``melting" topological defects at later epochs, as well as symmetry restoration~\cite{Tkachev:1995md, Greene:1997fu, Sato:2022jya} with very large initial amplitude of the dark Higgs, the present mechanism adds yet another pathway in Higgs portal dark sector models to generate cosmologically relevant topological defects and associated GWs. 
Thus if the symmetry is spontaneously broken in the zero temperature, which is the case for the axion or dark photon UV completions, and if the reheating temperature is high enough, the topogical defects and GWs may be very likely produced. Even if the symmetry is not broken like the case of WIMP, the topogical defects and GWs may be produced in the early Universe. 

The topological defects are subhorizon scale but long-lived, and they may induce the overdensities in the Universe forming miniclusters and primordial blackholes. Also if they remain until present Universe, These structures may contribute to cosmic birefringence~\cite{Agrawal:2019lkr, Takahashi:2020tqv, Yin:2021kmx, Kitajima:2022jzz, Gonzalez:2022mcx,Jain:2022jrp, Kitajima:2023kzu} and could also be probed by experiments searching for topological defects~\cite{Masia-Roig:2019hsy,GNOME:2023rpz} on Earth. 
In particular, the DWs can explain the hints from the isotropic cosmic birefringence~\cite{Minami:2020odp,Diego-Palazuelos:2022dsq}.

What I have found can be easily extend to the local $\U(1)$ symmetry or other larger symmetries, since the dynamics is mostly due to the dark Higgs evolution in the early Universe.   

\section*{Acknowledgments}

I thank my collaborators Diego Gonzalez, Fumiaki Kozai, Fuminobu Takahashi, Junseok Lee, Kai Murai, and Naoya Kitajima in a series of studies and ongoing projects \cite{Kitajima:2022jzz, Gonzalez:2022mcx, Kitajima:2023cek, Kitajima:2023kzu} for discussions on the lattice simulations, initial fluctuations, and GWs. I learned many during discussions in those projects. This work was supported by JSPS KAKENHI Grant Nos. 20H05851, 21K20364, 22K14029, 22H01215, and the Incentive Research Fund for Young Researchers from Tokyo Metropolitan University.
\clearpage
\appendix
\section{Setup for Lattice Simulation with Various Scenarios} 

Here I clarify the setups for the numerical lattice simulations in the main part and I also compare the other scnearios in this appendix to strenghen the main result. The setup and definitions of the scenarios are shown in Table~\ref{tab:cosmo_scenarios}. For each setup, I consider the case of a real Higgs for the $Z_2$ model and a complex Higgs for the global $ U(1)$ model.

The scenarios have the following meanings as indicated in the table: scenario (SB) denotes the usual symmetry-breaking scenario with a constant mass. (thSB) denotes the conventional melting topological defect scenarios, where the symmetry is preserved initially with the temperature-dependent negative mass squared. (NSB) is the usual non-symmetry-breaking scenario with a constant mass, i.e., initially we place the field away from the symmetry-enhanced point. (thNSB) is the scenario studied in the main part, where the negative thermal masses dominate the system but initially the symmetry is broken. So far, the scenario is all in the radiation-dominated Universe. (thNSBmat) denotes the scenario of (thNSB) but assuming a matter-dominated Universe. Note that this is not the scenario where the thermal mass is induced by the diluting plasma, $T \propto a^{-3/8}$, which roughly corresponds to (NSB) since the temperature does not change much. Rather, this corresponds to the case where the temperature is much higher than the diluting plasma one, which is possible if some other long-lived particle, such as moduli, dominate the Universe and decay late. Alternatively, we may consider this as the usual matter-dominated epoch around the present Universe, which may be important for the scenarios searching for topological defects on Earth or cosmic-birefringences.

I will show the results for different fluctuations and types of models one by one as follows.

\begin{table}[h]
    \centering
    \caption{Cosmological Scenarios and Initial Fluctuations for Simulation. In all scenarios, the momentum of the initial fluctuation is cutoff at the initial Hubble parameter $H_0 = d$(Thus I do not perform the renormalization due to the large fluctuations~\cite{Berges:2013lsa}.). The initial conformal time is taken to be $\tau_0 = 1$. In the radiation-dominated Universe, we have $a = \tau d=1$. The infrared cutoff is $a k_{\rm IR} = a \times 2\pi/L = 0.02d$. Here $d$ is the machine unit, which can be taken arbitrarily without changing the results. EoS denotes the equation of state.}
    \label{tab:cosmo_scenarios}
    \renewcommand{\arraystretch}{1.2} % Adjust row height
    \setlength{\tabcolsep}{0.5pt} % Adjust column separation
    \rowcolors{2}{gray!10}{white} % Shade even rows
    \begin{tabular}{@{}p{3cm}cc>{\centering\arraybackslash}p{1.5cm}@{}} % Adjust column widths
        \toprule
        \textbf{Scenario} & $\bar{\phi}_i^0$ & $m/H_0$ & {EoS}  \\
        \midrule
        SB & $0$ & $0.064$ & $1/3$  \\
        thSB & $0$ & $0.08/a$ & $1/3$  \\
        NSB & $0.1 \delta_{i1}$ & $0.064$ & $1/3$ \\
        \textbf{thNSB(main part)} & $0.1 \delta_{i1}$ & $0.08/a$ & $1/3$ \\
        thNSBmat & $0.1 \delta_{i1}$ & $0.08/a$ & $0$ \\
        \bottomrule
        \multicolumn{4}{l}{\textbf{Initial Fluctuation}} \\
        \midrule
        \multicolumn{4}{p{0.371\linewidth}}{\small Gaussian fluctuation: $c = 0.01/d$} \\
        \multicolumn{4}{p{0.371\linewidth}}{\small Scale-invariant fluctuation: $H^2_{\rm inf} = 0.002d^2$} \\
        \bottomrule
    \end{tabular}
\end{table}

To study the GW, I define\footnote{This is not $\Omega_{\rm GW}$ from the output of {\tt CosmoLattice}. In the output, it is defined by $\frac{d}{d \log k}\rho_{\rm GW}/\rho_{\rm tot}$ with $\rho_{\rm tot}$ being the total dynamical energy density in the estimation in the fixed background option rather than the critical density.}
\beq
\Omega_{\rm gw} \equiv \frac{a^4 \times \frac{d}{d \log k} \rho_{\rm GW}}{3 M_{\rm pl}^2 H_0^2},
\eeq
where $\rho_{\rm GW}$ is the energy density of the GWs, and $a$ is normalized to unity at the initial time. I also emphasize that to obtain $\Omega_{\rm GW}^0$ in the present Universe, we need to take into account the redshift as well as entropy dilution \cite{Kitajima:2023cek}. 

In each section below, I show the snapshot at $\tau = 151/d$, and the GW spectrum and reduced power spectra of the scalar fields, which is knowe to provide valuable information~\cite{Kitajima:2022jzz}. The snapshot is taken to be the contour of $\O(1\%)$ of the small value in the histogram of $|\F|$. 
Here I denote some common observations. 

In all cases, I found from the snapshots that the temperature-dependent potential is essential to have the topological defects even if the symmetry is always broken, which can be seen by comparing the (thNSB) and (NSB) scenarios. For comparison, the mass becomes comparable to the Hubble parameter around $\tau = \mathcal{O}(10)$ in all cases. Additionally, I can also observe small topological defects formed in the (thNSB) and (thNSBmat) scenarios. In (SB) and (thSB), which are the melting topological defect scenarios \cite{Emond:2021vts, Babichev:2021uvl, Babichev:2023pbf, Dankovsky:2024ipq}, we have both small-sized and large-sized defects over the horizon.

 The GWs gets additional contribution from the peaks which can be cleary seen in the 
(thNSBmat) cases, which may be due to the contribution from the self-resonance and the fragmentation of the condensate (see e.g. Refs.\,~\cite{ Lozanov:2017hjm,Lozanov:2019ylm, Chatrchyan:2020pzh, Garcia:2023eol}).

I mention that although the Minkowski fluctuation and thermal fluctuation of melting DWs at the massless limit were carefully studied in \cite{Dankovsky:2024ipq}, the numerical simulations for the melting DWs and CSs with the gaussian and inflationary fluctuations were not performed before.

\subsection{$Z_2$ Models with Gaussian Fluctuations}

\begin{figure}[!h] 
    \begin{center}
    \includegraphics[width=160mm]{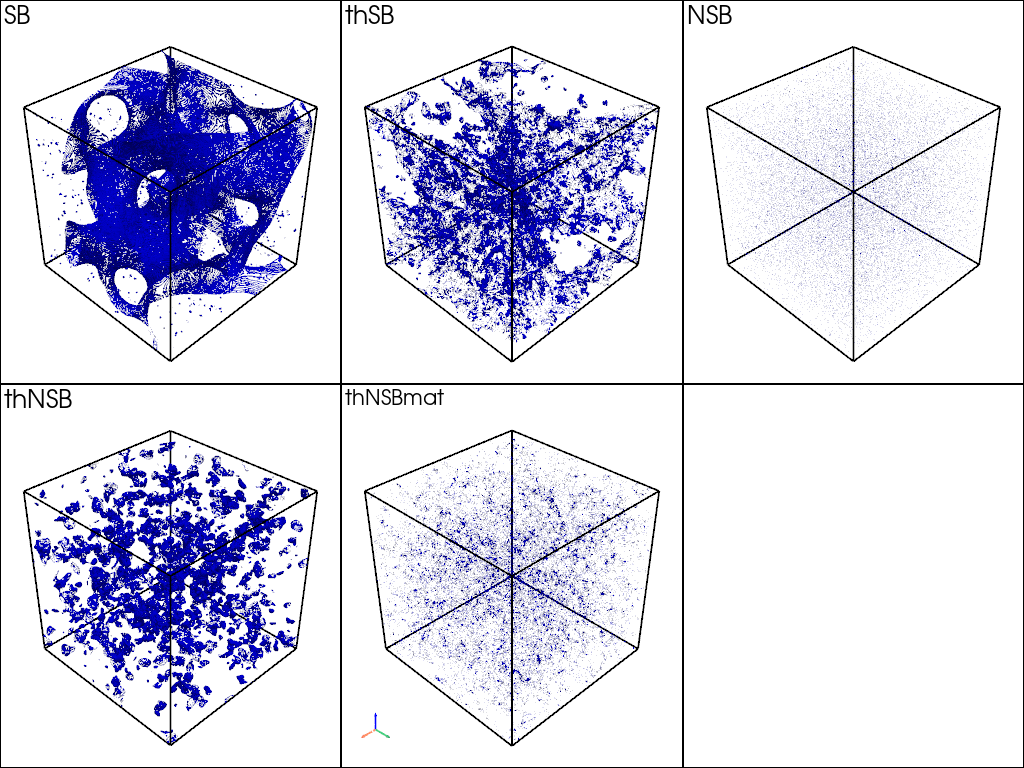}
    \end{center} 
    \caption{Snapshot of $Z_2$ models with Gaussian fluctuations at $\tau = 151/d$.}
    \label{fig:Z2gaussian} 
\end{figure}

\begin{figure}[!h] 
    \begin{center}
    \includegraphics[width=160mm]{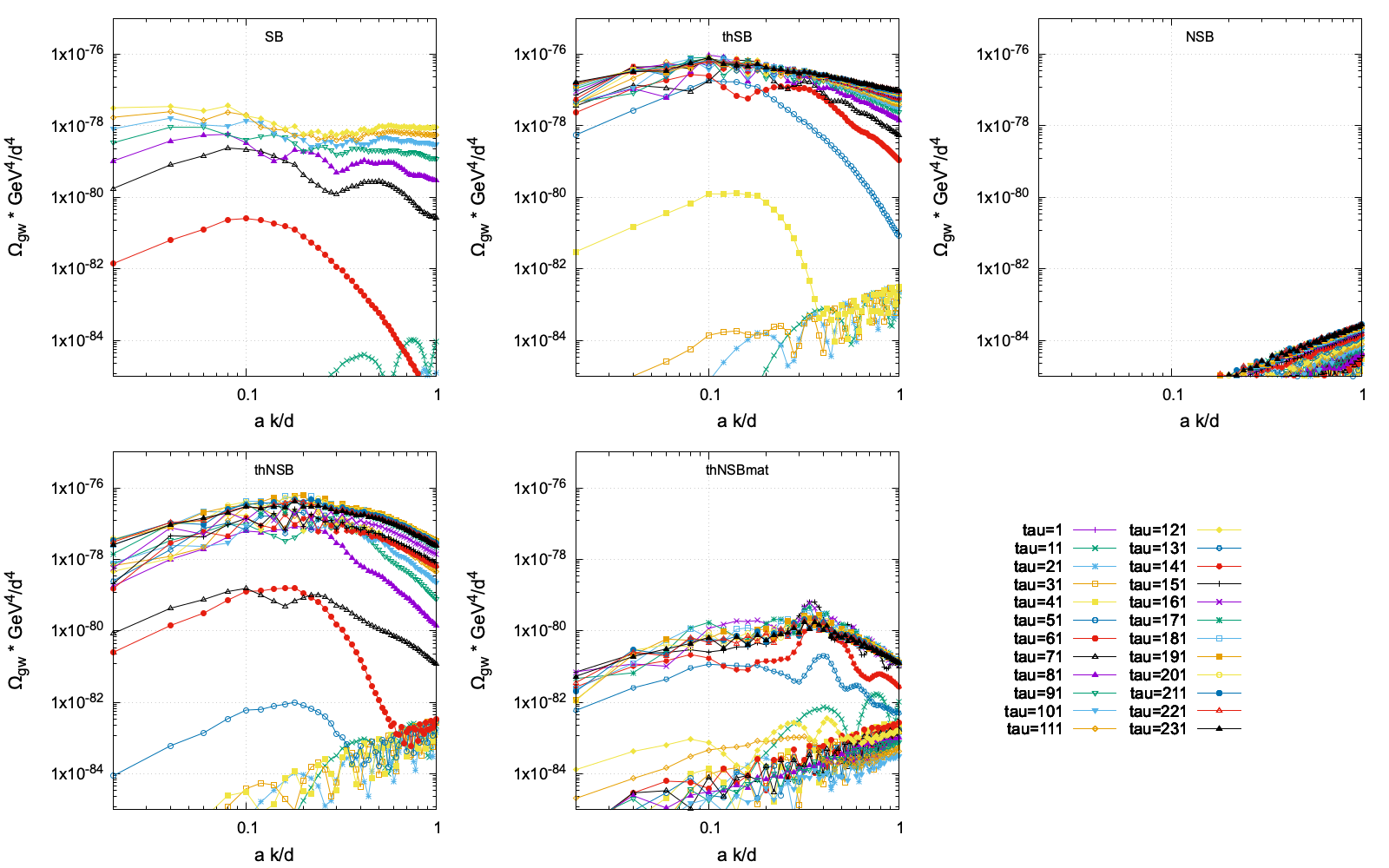}
    \end{center} 
    \caption{GW spectra from $Z_2$ models with Gaussian fluctuations.}
    \label{fig:Z2gaussianGW} 
\end{figure}

\begin{figure}[!h] 
    \begin{center}
    \includegraphics[width=160mm]{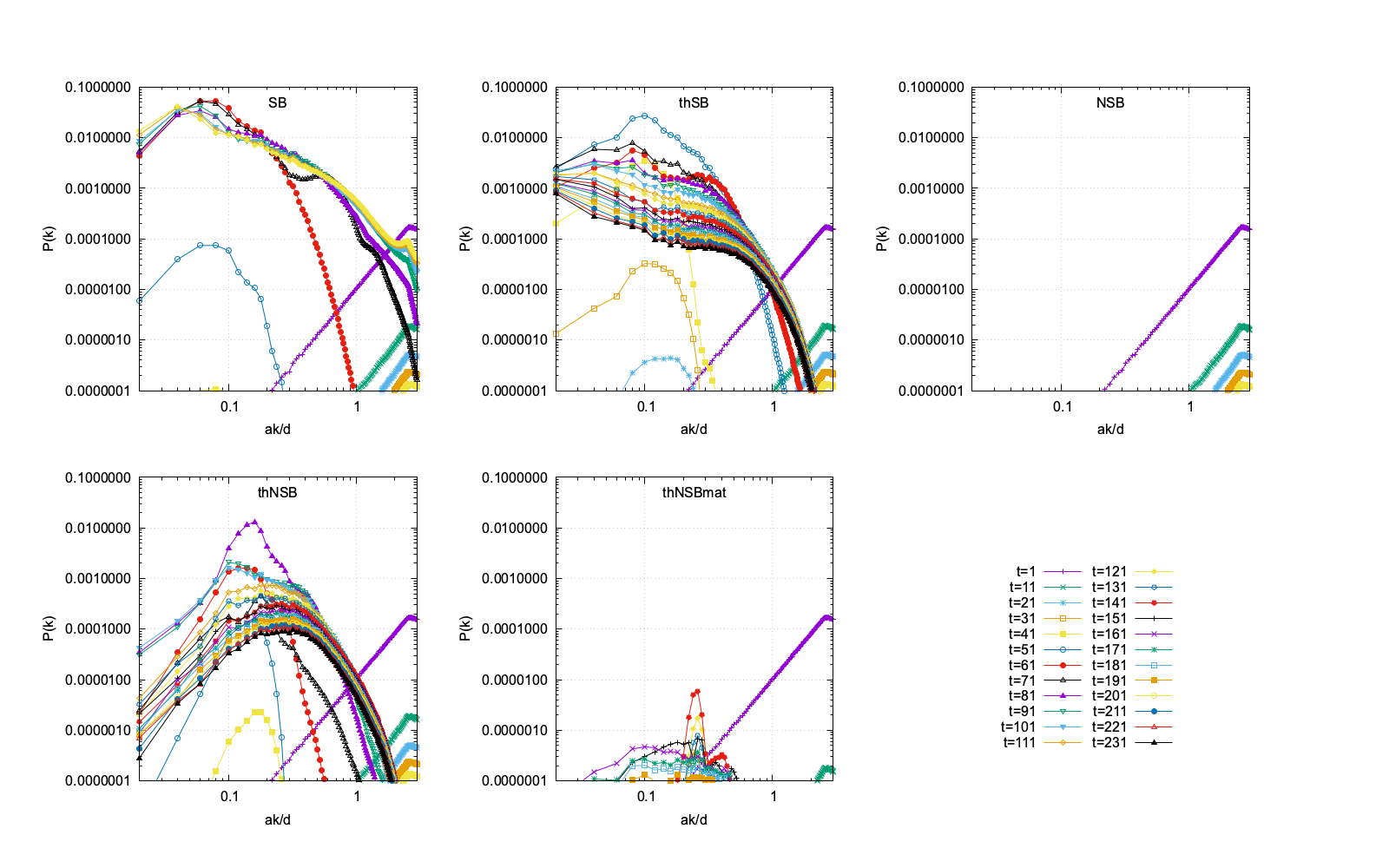}
    \end{center} 
    \caption{Evolution of the reduced power spectrum of $\phi$ in the $Z_2$ model with Gaussian fluctuations.}
    \label{fig:Z2gaussianspectra} 
\end{figure}

\clearpage

\subsection{$Z_2$ Models with Scale-Invariant Fluctuations}

\begin{figure}[!h] 
    \begin{center}
    \includegraphics[width=160mm]{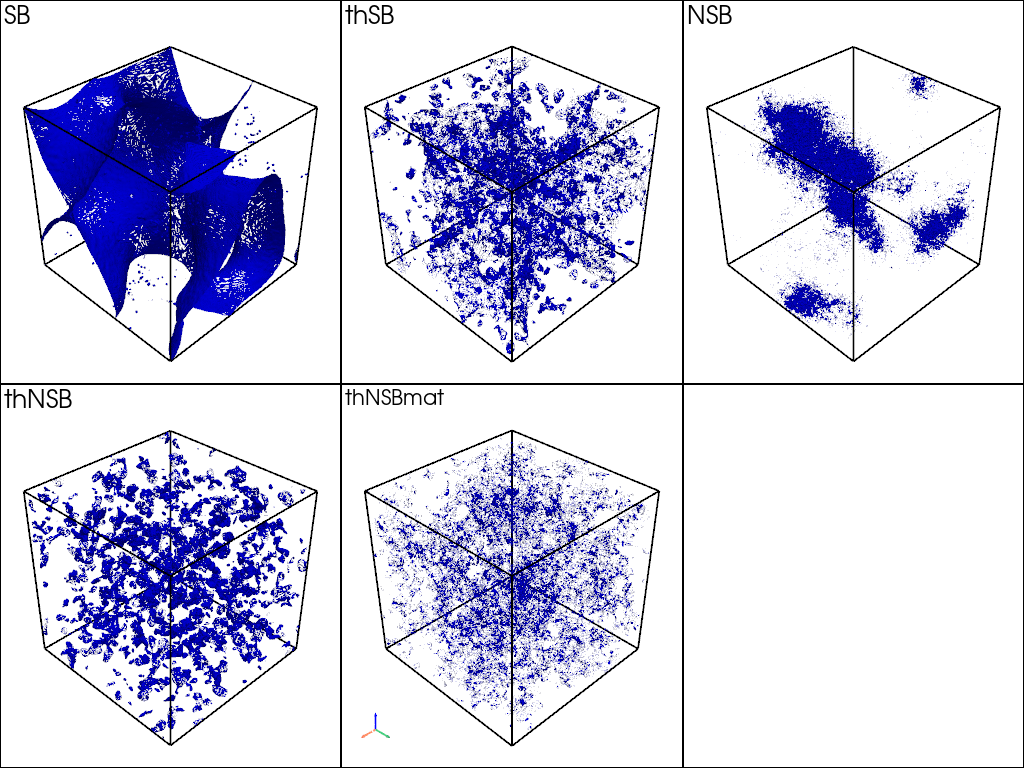}
    \end{center} 
    \caption{Snapshot of $Z_2$ models with scale-invariant fluctuations at $\tau = 151/d$.}
    \label{fig:Z2scaleinv} 
\end{figure}

\begin{figure}[!h] 
    \begin{center}
    \includegraphics[width=160mm]{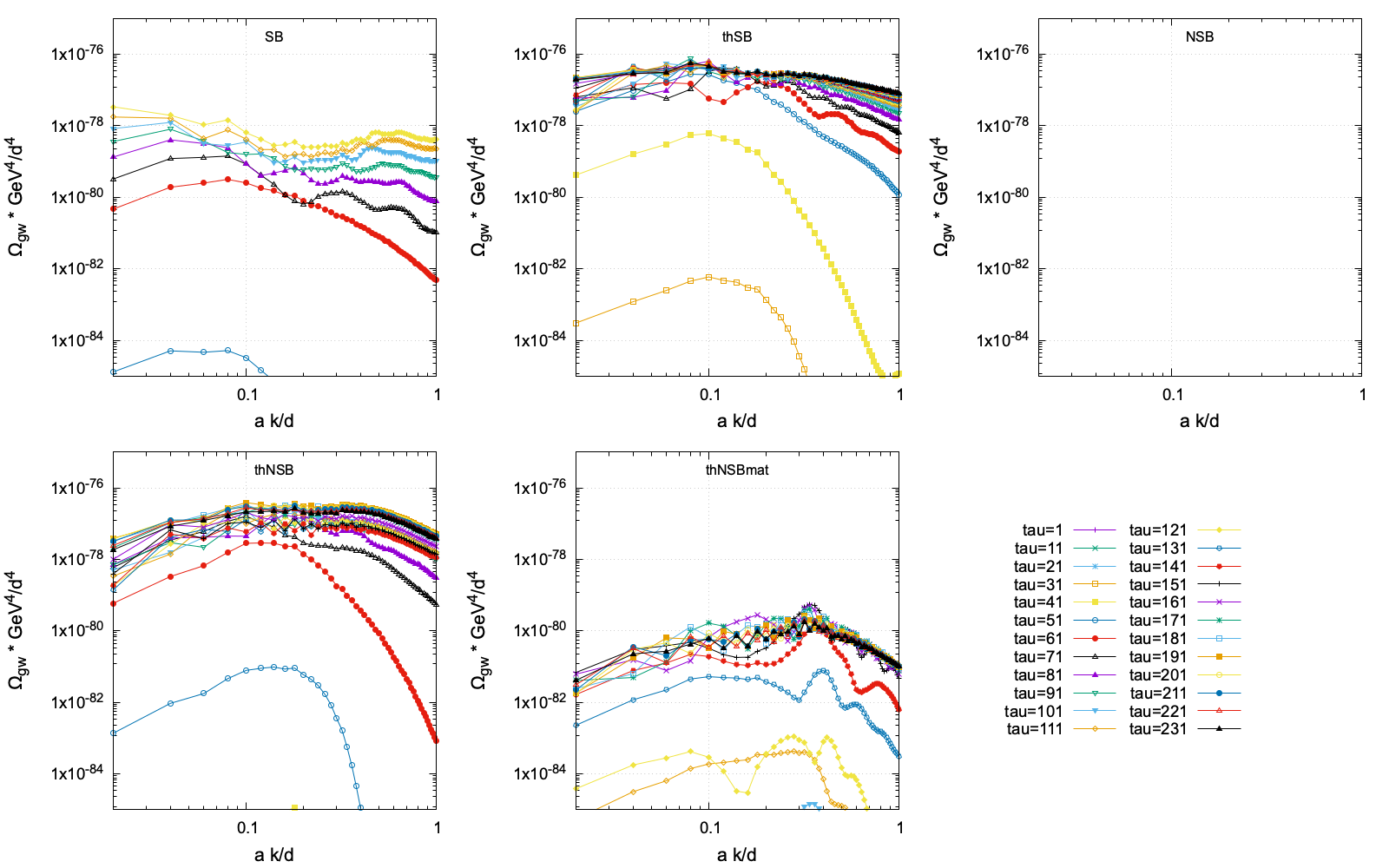}
    \end{center} 
    \caption{GW spectra from $Z_2$ models with scale-invariant fluctuations.}
    \label{fig:Z2scaleinvGW} 
\end{figure}

\begin{figure}[!h] 
    \begin{center}
    \includegraphics[width=160mm]{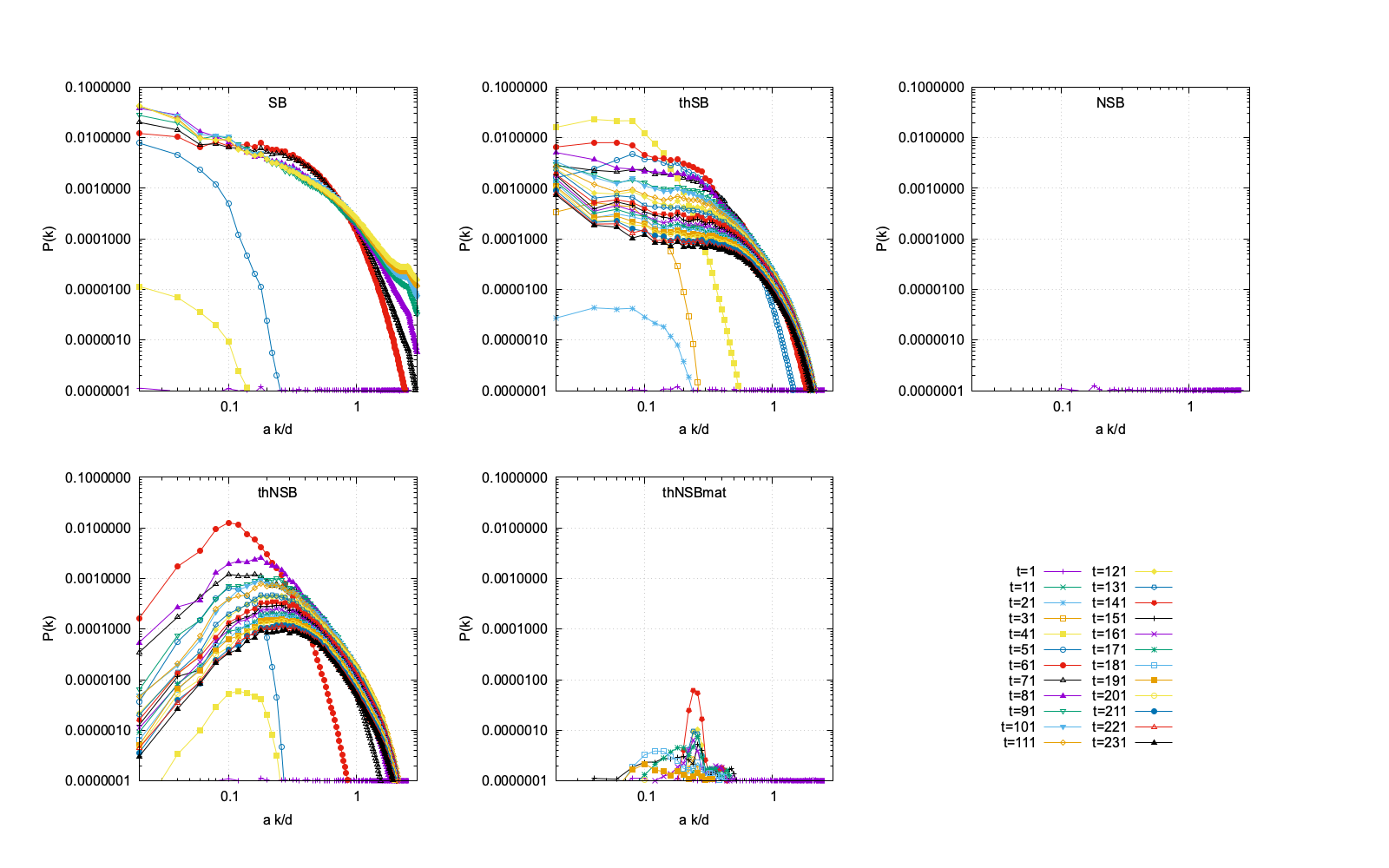}
    \end{center} 
    \caption{Evolution of the reduced power spectrum of $\phi$ in the $Z_2$ model with scale-invariant fluctuations.}
    \label{fig:Z2scaleinvspectra} 
\end{figure}

\clearpage

\subsection{$U(1)$ Models with Gaussian Fluctuations}

\begin{figure}[!h] 
    \begin{center}
    \includegraphics[width=160mm]{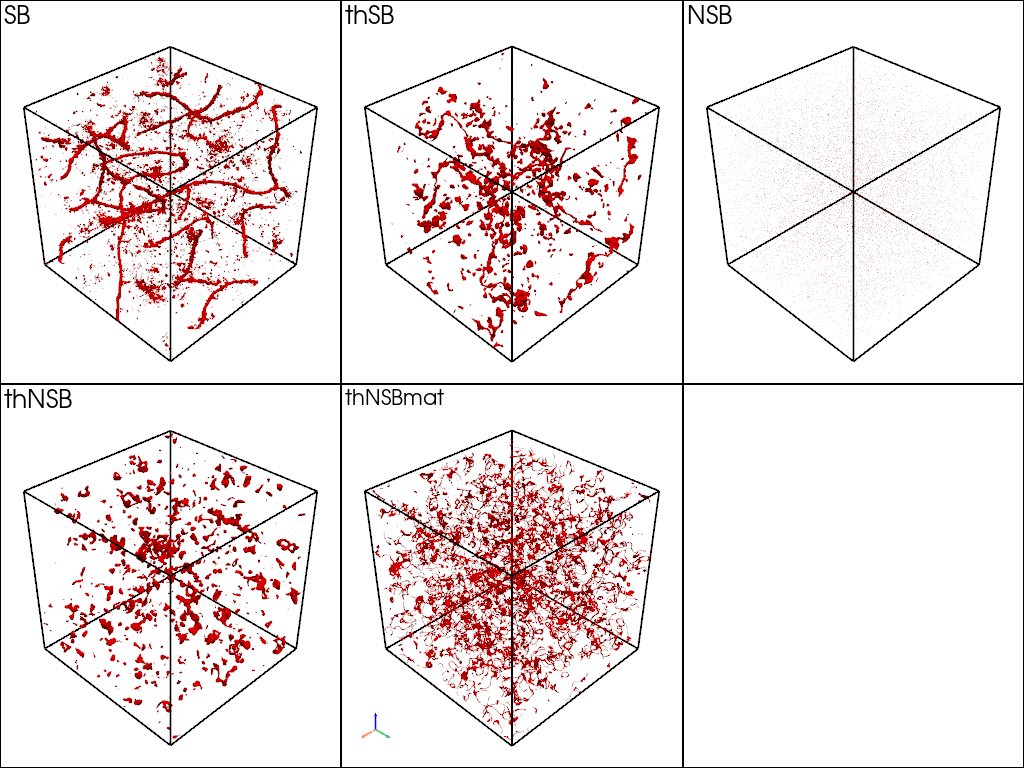}
    \end{center} 
    \caption{Snapshot of $U(1)$ models with Gaussian fluctuations at $\tau = 151/d$.}
    \label{fig:U1gaussian} 
\end{figure}

\begin{figure}[!t] 
    \begin{center}
    \includegraphics[width=160mm]{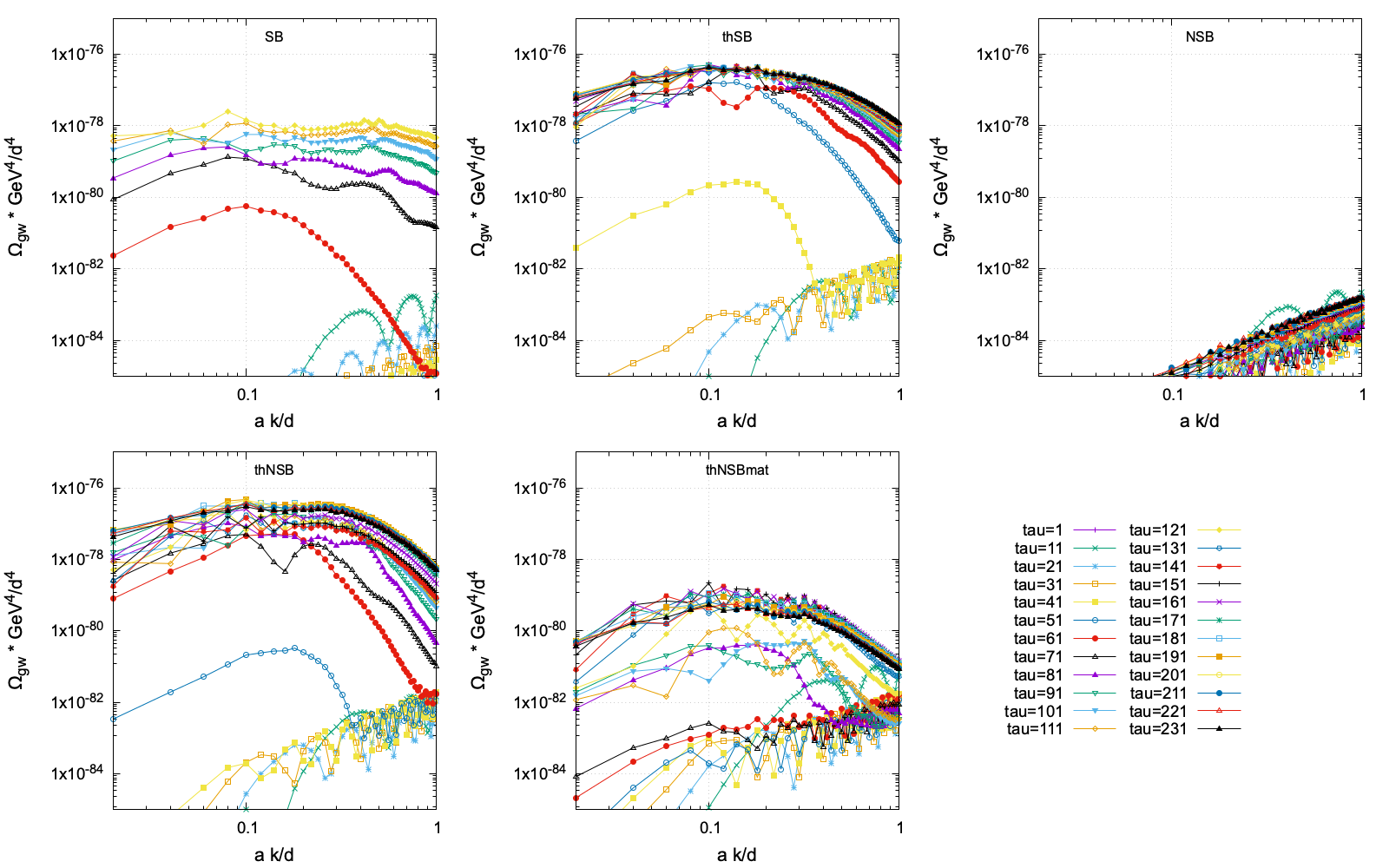}
    \end{center} 
    \caption{GW spectra from $U(1)$ models with Gaussian fluctuations.}
    \label{fig:U1gaussianGW} 
\end{figure}

\begin{figure}[!h] 
    \begin{center}
    \includegraphics[width=150mm]{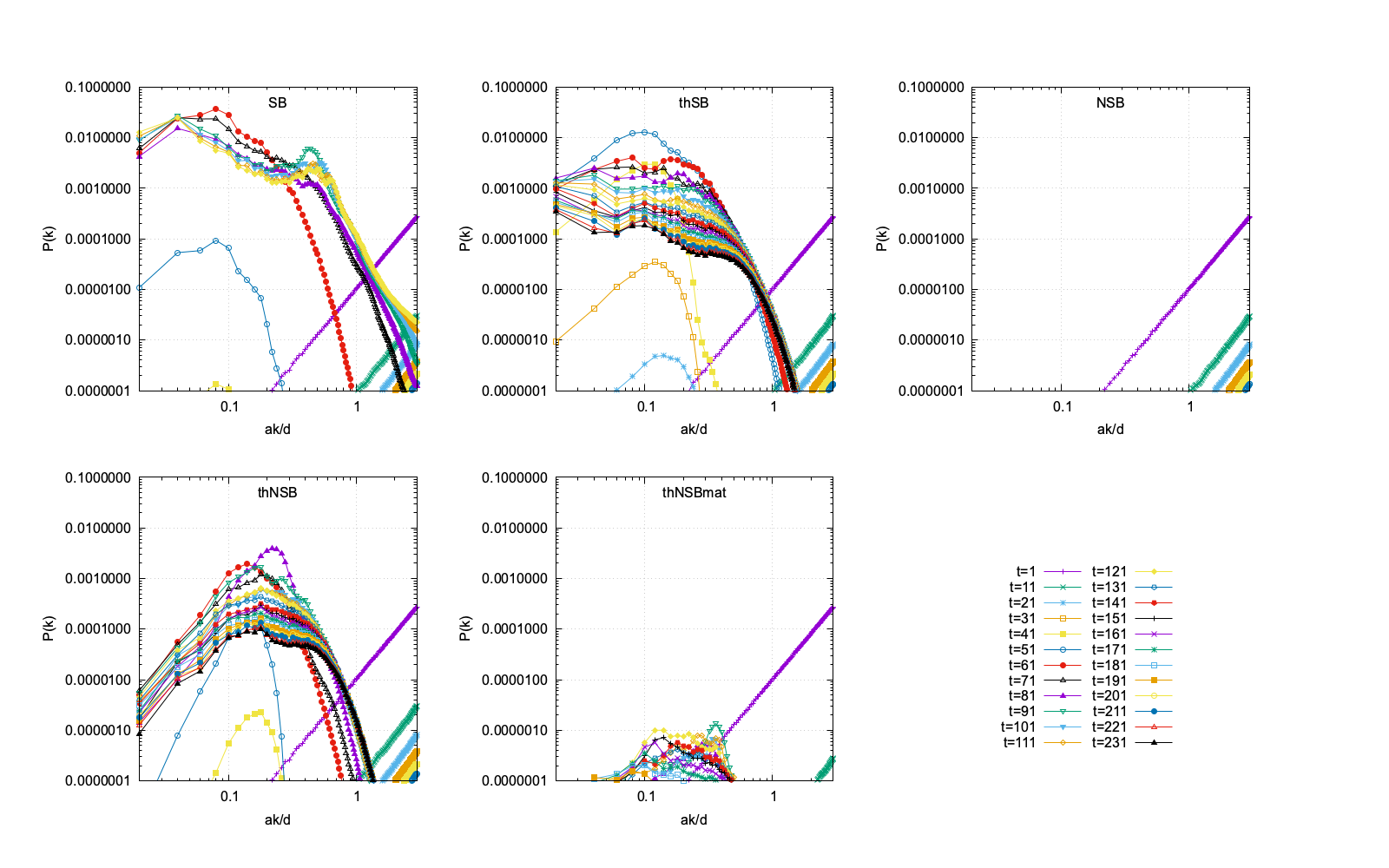}
    \includegraphics[width=150mm]{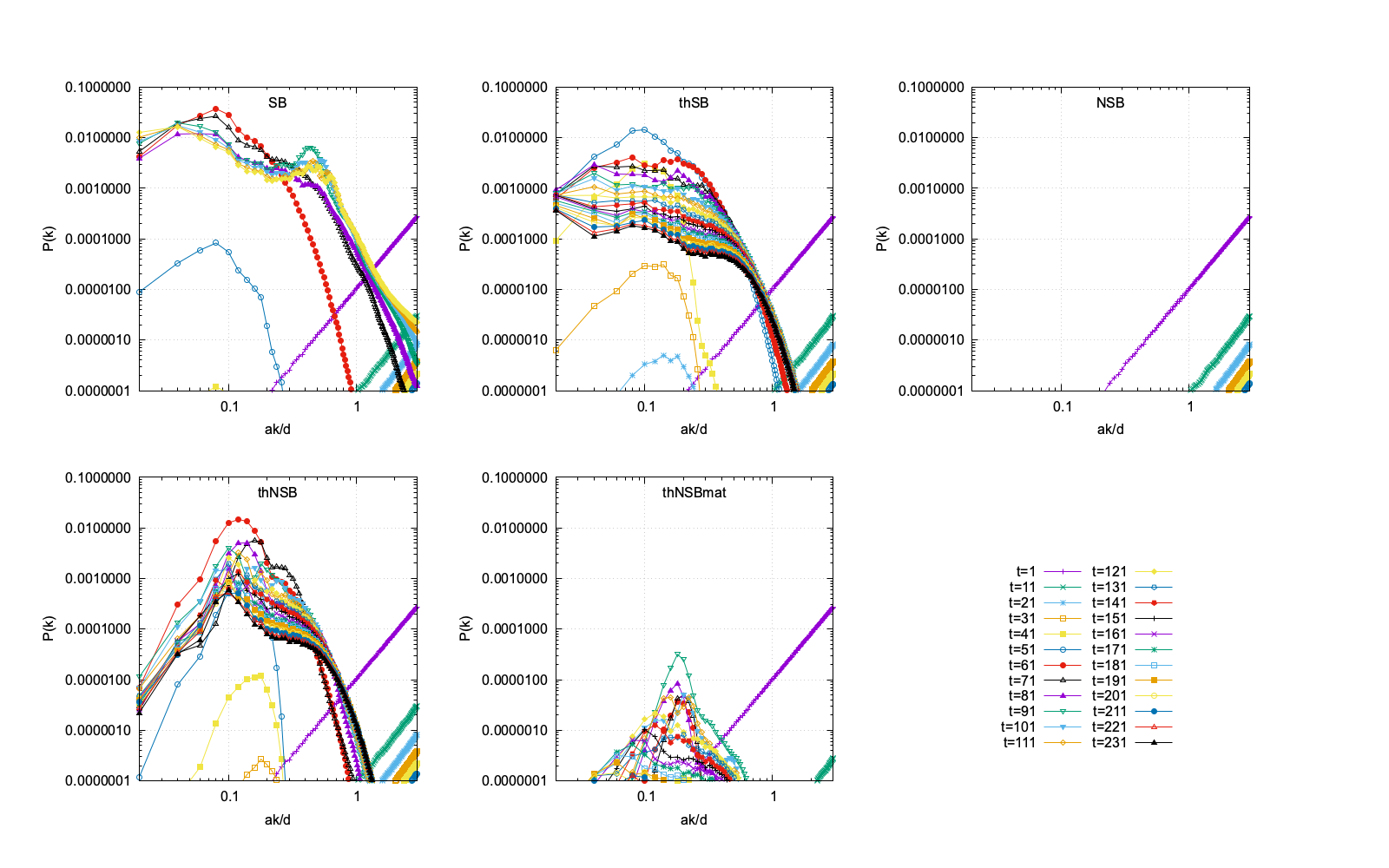}
    \end{center} 
    \caption{Evolution of the reduced power spectrum of $\phi_1$ (top) and $\phi_2$ (bottom) in the $U(1)$ model with Gaussian fluctuations.}
    \label{fig:U1gaussianspectra} 
\end{figure}

\clearpage

\subsection{$U(1)$ Models with Scale-Invariant Fluctuations}

\begin{figure}[!h] 
    \begin{center}
    \includegraphics[width=160mm]{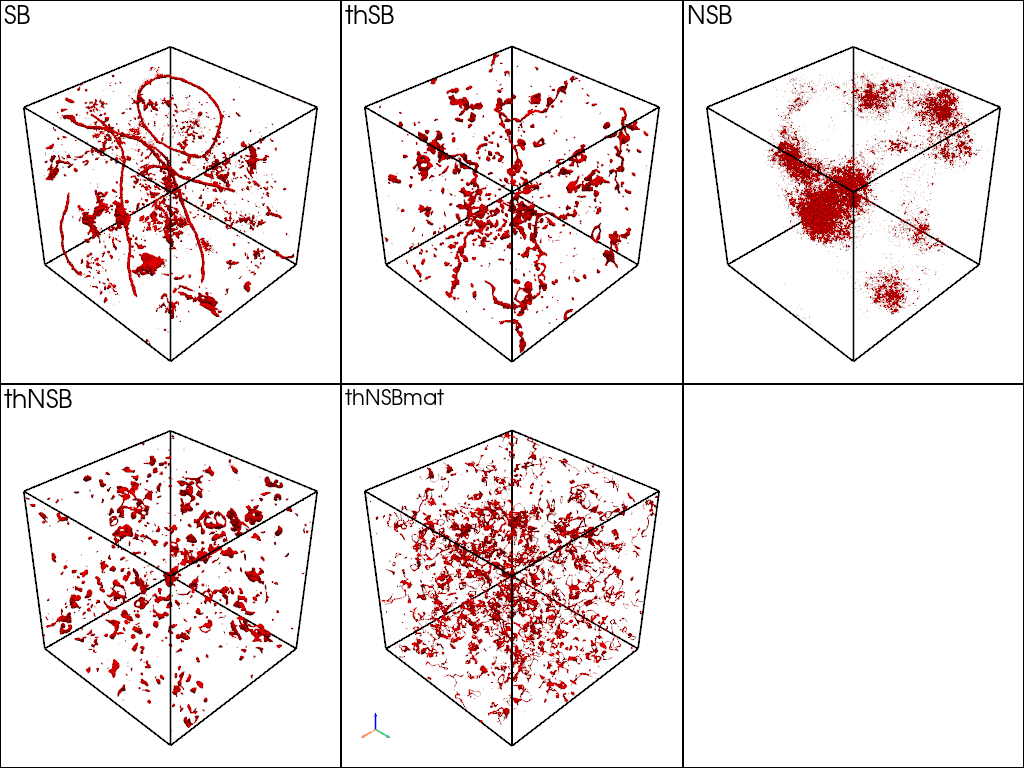}
    \end{center} 
    \caption{Snapshot of $U(1)$ models with scale-invariant fluctuations at $\tau = 151/d$. $H_{\rm inf}^2 = 0.01d^2$.}
    \label{fig:U1scaleinv_large} 
\end{figure}

\begin{figure}[!t] 
    \begin{center}
    \includegraphics[width=160mm]{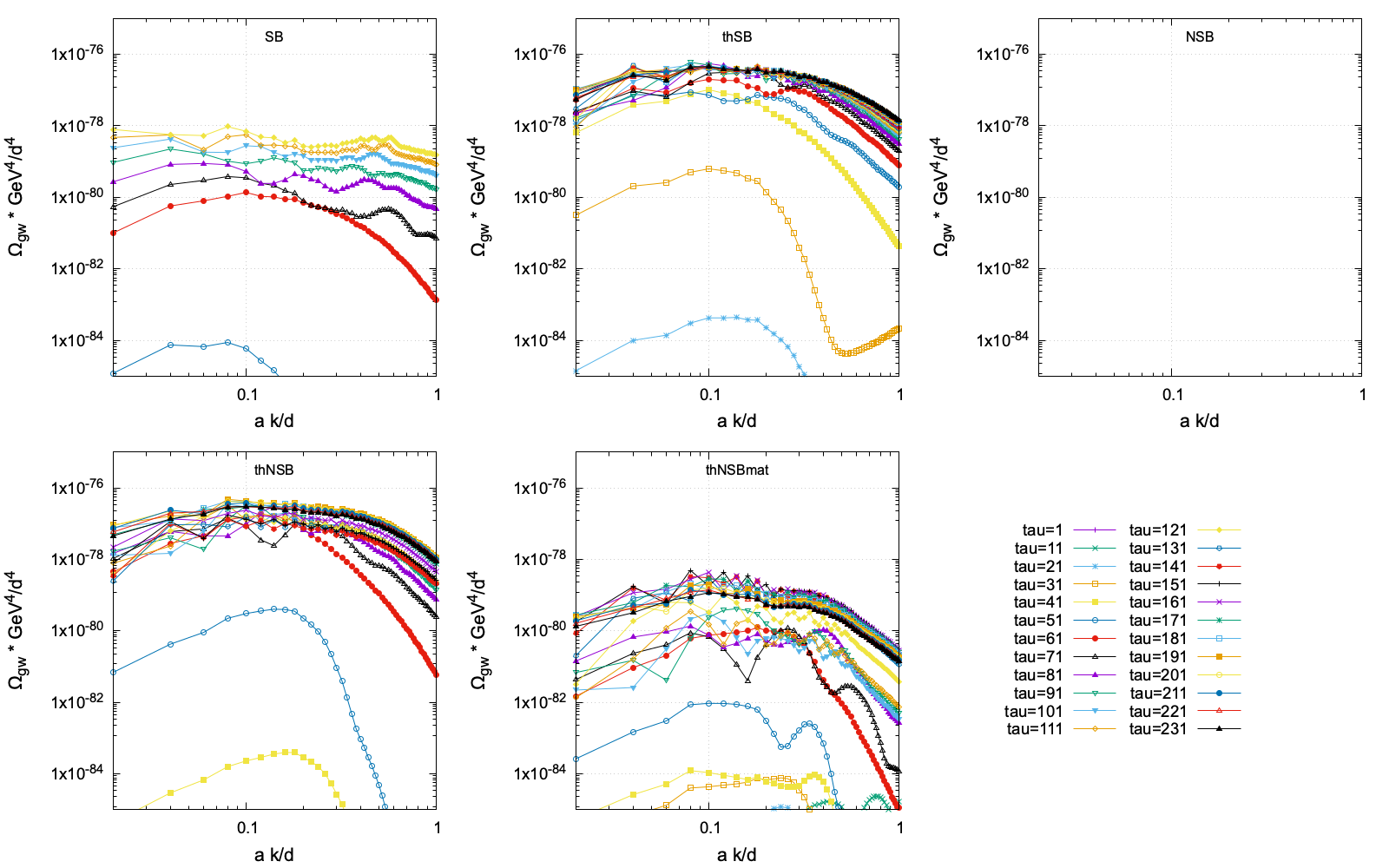}
    \end{center} 
    \caption{GW spectra from $U(1)$ models with scale-invariant fluctuations. $H_{\rm inf}^2 = 0.01d^2$.}
    \label{fig:U1scaleinvGWlarge} 
\end{figure}

\begin{figure}[!t] 
    \begin{center}
    \includegraphics[width=160mm]{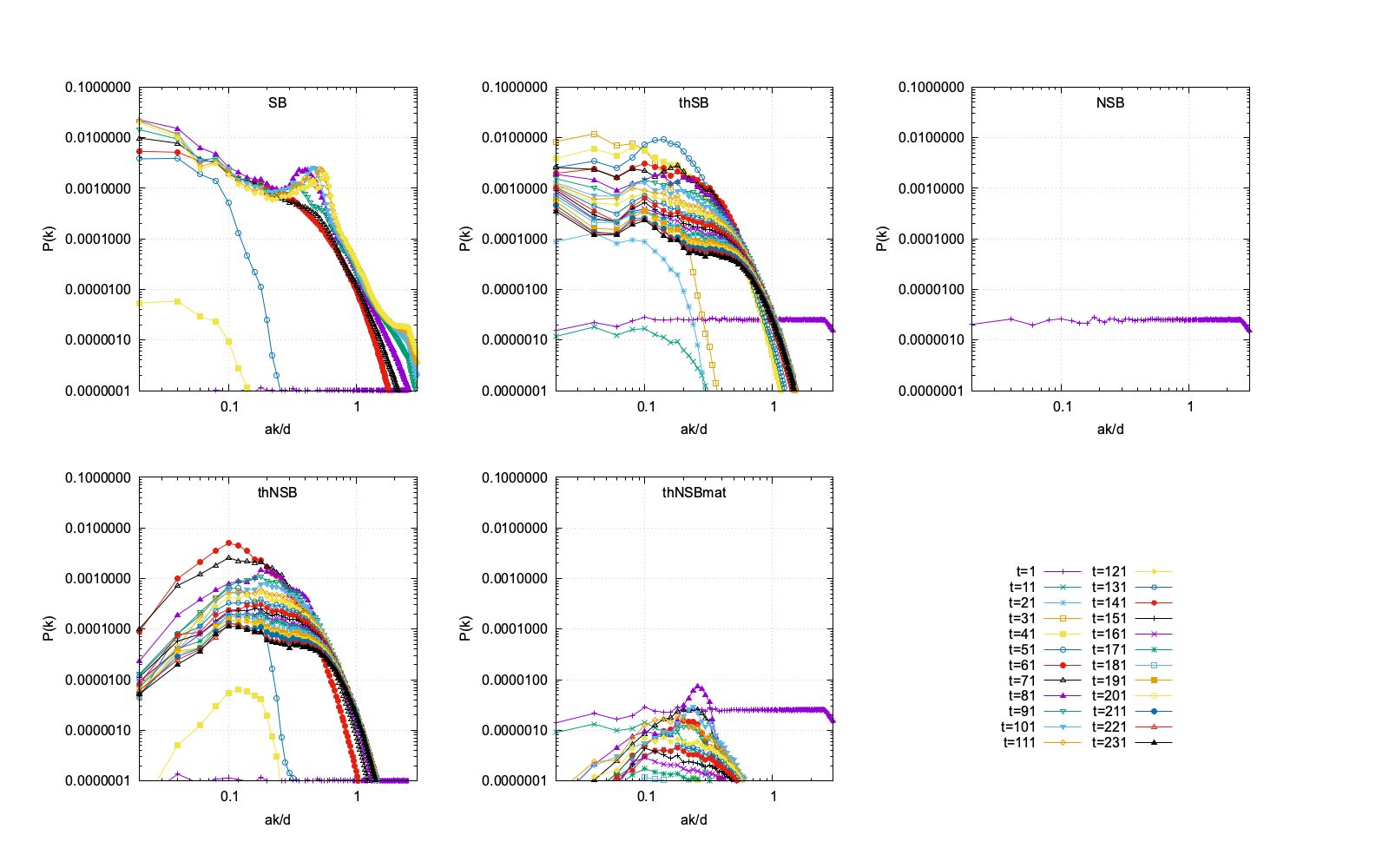}
    \includegraphics[width=160mm]{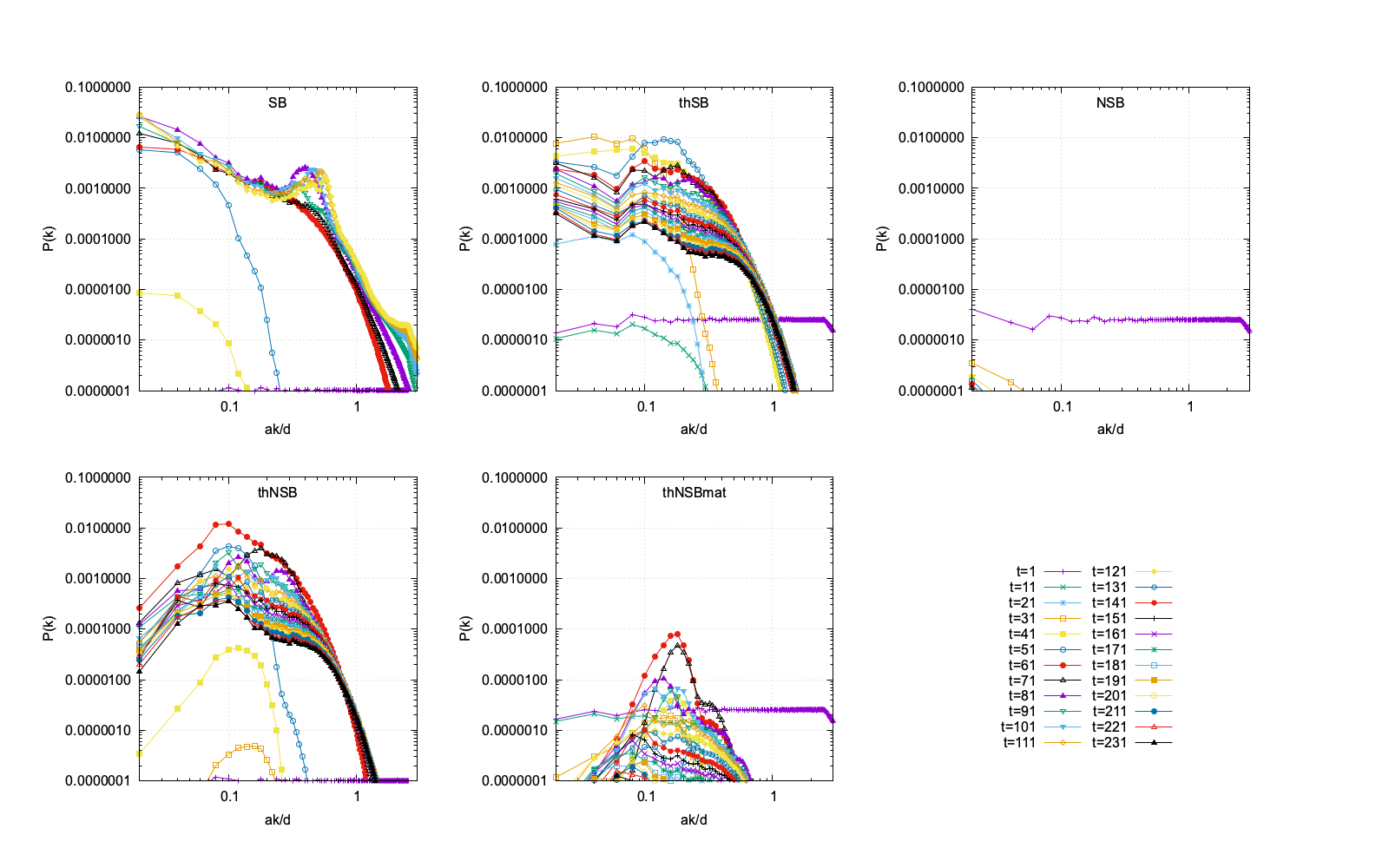}
    \end{center} 
    \caption{Reduced power spectrum of $\phi_1$ (top panel) and $\phi_2$ (bottom panel) in $U(1)$ models with scale-invariant fluctuations. $H_{\rm inf}^2 = 0.01d^2$.}
    \label{fig:U1scaleinvpslarge} 
\end{figure}

\clearpage

\section{Larger Scale-Invariant Fluctuations}

It was shown that inflationary fluctuations or scale-invariant fluctuations can make the topological defects stable against population bias in \cite{Gonzalez:2022mcx,Kitajima:2023kzu}. This feature also exists in the non-symmetry restoration case with the negative thermal mass squared. To see this, I set $H_{\rm inf}^2 = 0.01d^2$, which is larger than the choices in the main discussion. In the (thNSB) scenario, the topological defects seem to be rather larger than before, which essentially confirms the claim in \cite{Gonzalez:2022mcx,Kitajima:2023kzu}. For the (NSB) scenario, no topological defects remain, which implies that the enhancement of fluctuations due to the negative thermal potential may make the fluctuations enhanced and thus the stability is more robust. 

\subsection{$Z_2$ Case}

\begin{figure}[!t] 
    \begin{center}
    \includegraphics[width=160mm]{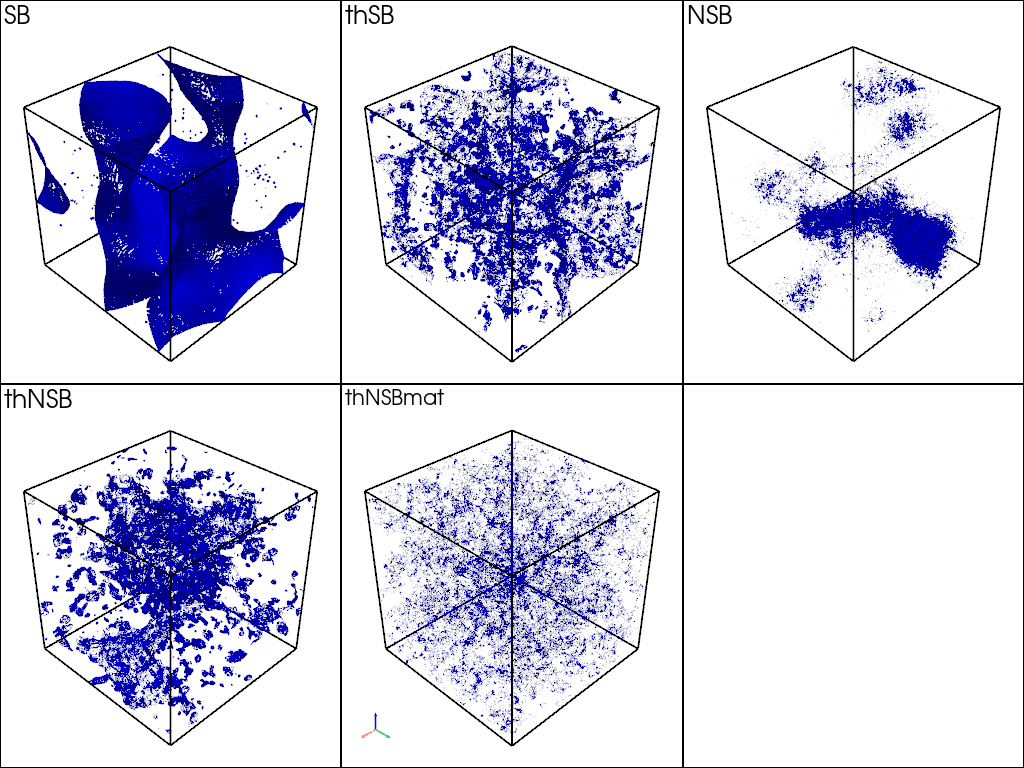}
    \end{center} 
    \caption{Snapshot of $Z_2$ models with scale-invariant fluctuations at $\tau = 151/d$. $H_{\rm inf}^2 = 0.01d^2$.}
    \label{fig:Z2scaleinvGWlarge} 
\end{figure}

\begin{figure}[!t] 
    \begin{center}
    \includegraphics[width=160mm]{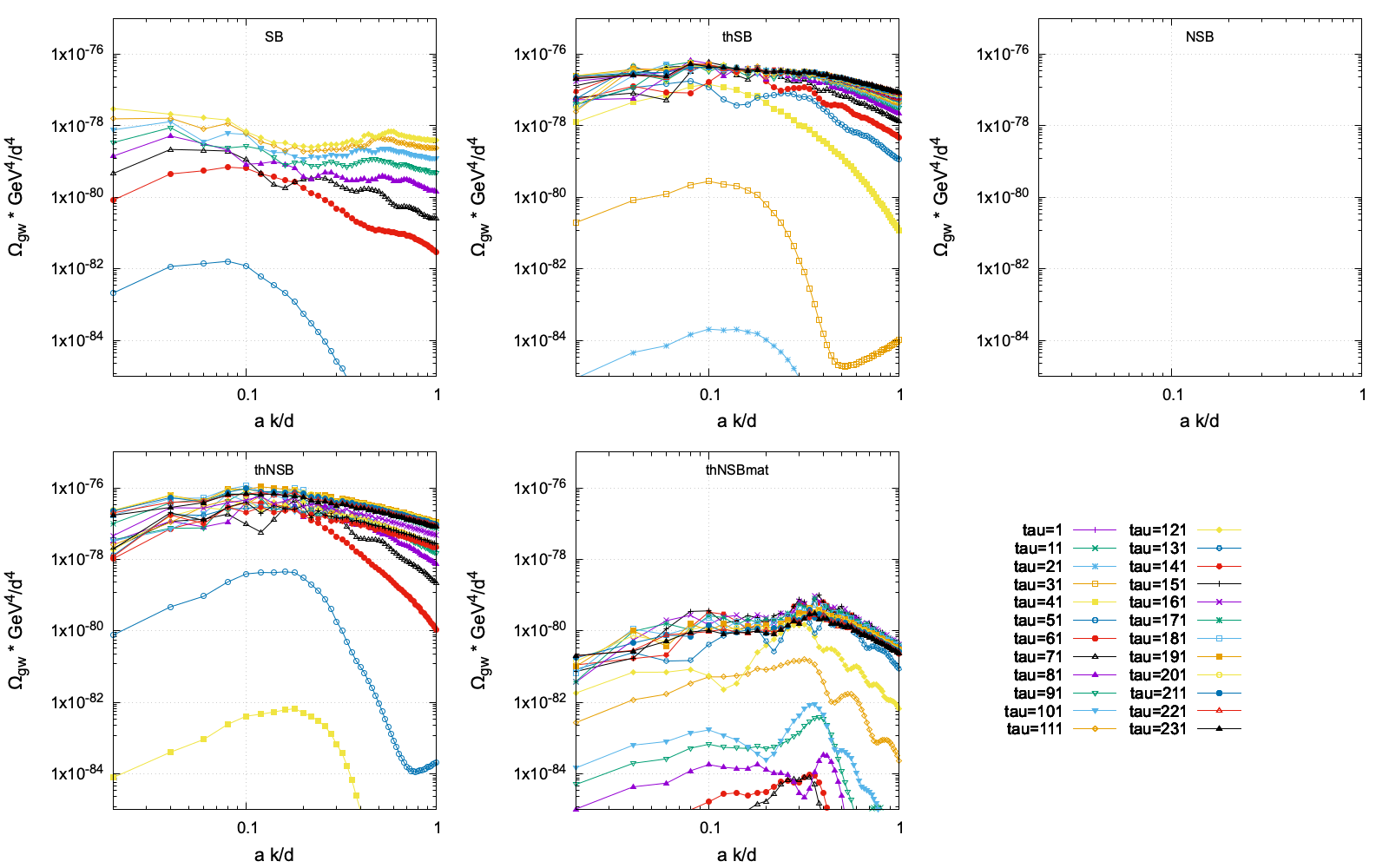}
    \end{center} 
    \caption{GW spectra from $Z_2$ models with scale-invariant fluctuations. $H_{\rm inf}^2 = 0.01d^2$.}
    \label{fig:Z2scaleinvGWlarge} 
\end{figure}

\begin{figure}[!t] 
    \begin{center}
    \includegraphics[width=160mm]{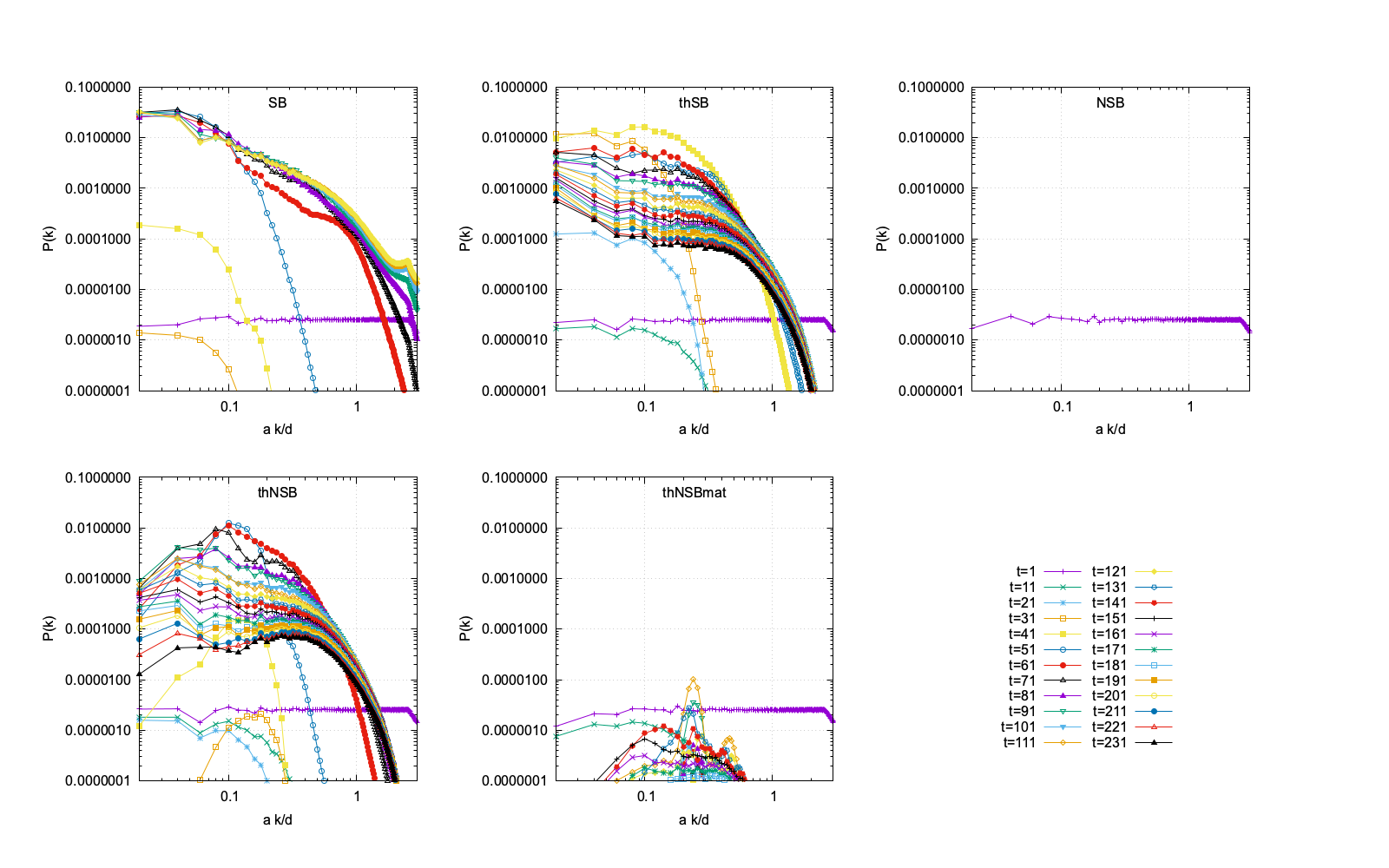}
    \end{center} 
    \caption{Reduced power spectrum of $\phi$ in $Z_2$ models with scale-invariant fluctuations. $H_{\rm inf}^2 = 0.01d^2$.}
    \label{fig:Z2scaleinvspectra_large} 
\end{figure}
\clearpage

\subsection{$U(1)$ Case}

\begin{figure}[!h] 
    \begin{center}
    \includegraphics[width=160mm]{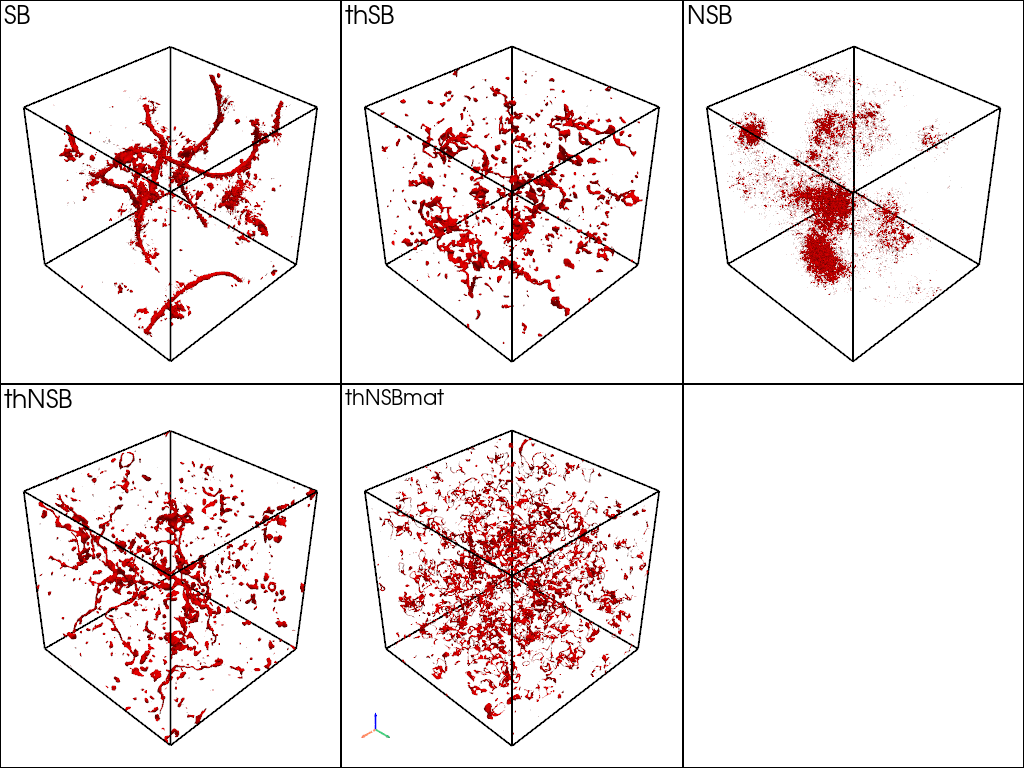}
    \end{center} 
    \caption{Snapshot of $U(1)$ models with scale-invariant fluctuations at $\tau = 151/d$. $H_{\rm inf}^2 = 0.01d^2$.}
    \label{fig:U1scaleinvGWlarge_snap} 
\end{figure}

\begin{figure}[!h] 
    \begin{center}
    \includegraphics[width=160mm]{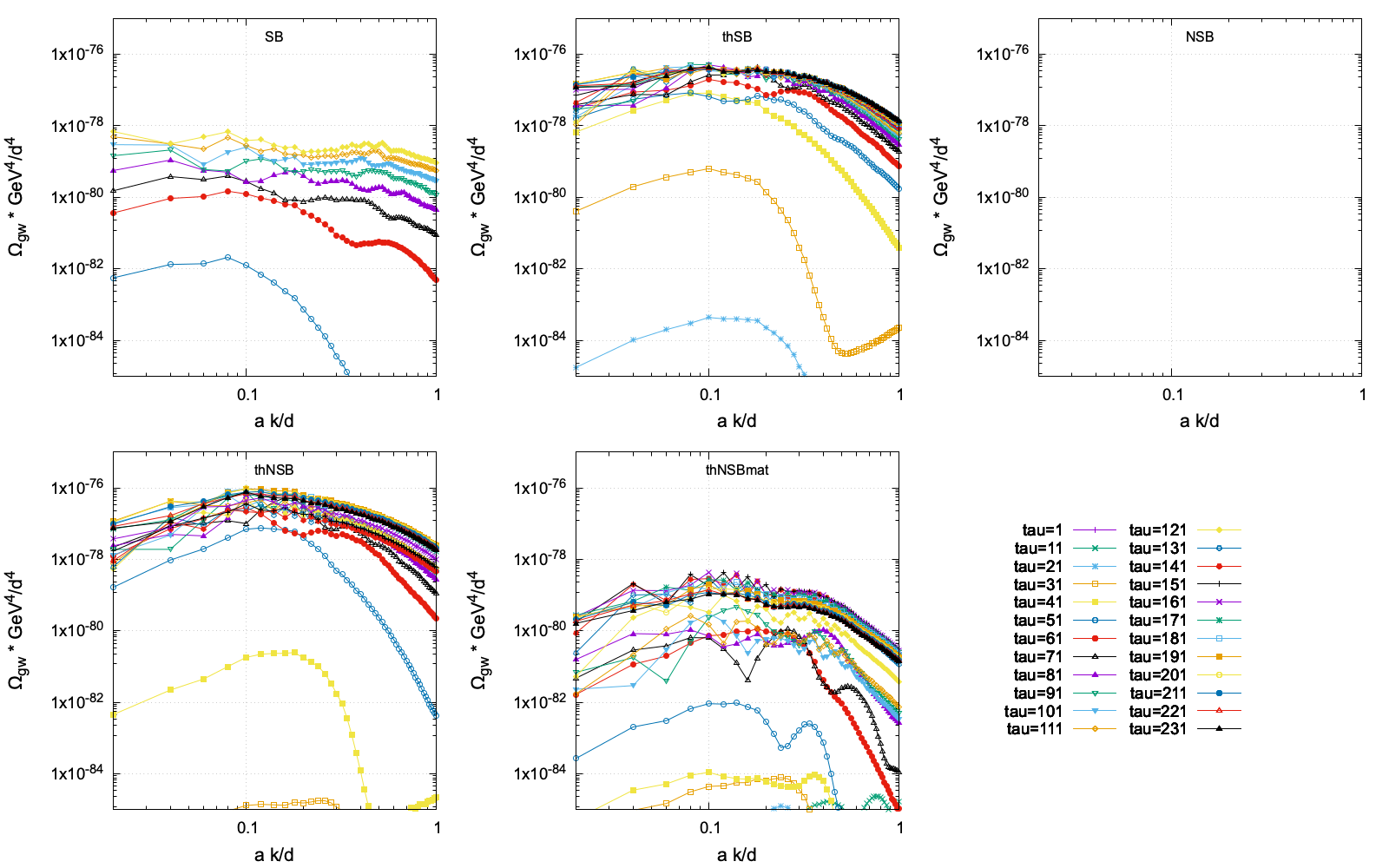}
    \end{center} 
    \caption{GW spectra from $U(1)$ models with scale-invariant fluctuations. $H_{\rm inf}^2 = 0.01d^2$.}
    \label{fig:U1scaleinvGWlarge} 
\end{figure}

\begin{figure}[!h] 
    \begin{center}
\includegraphics[width=160mm]{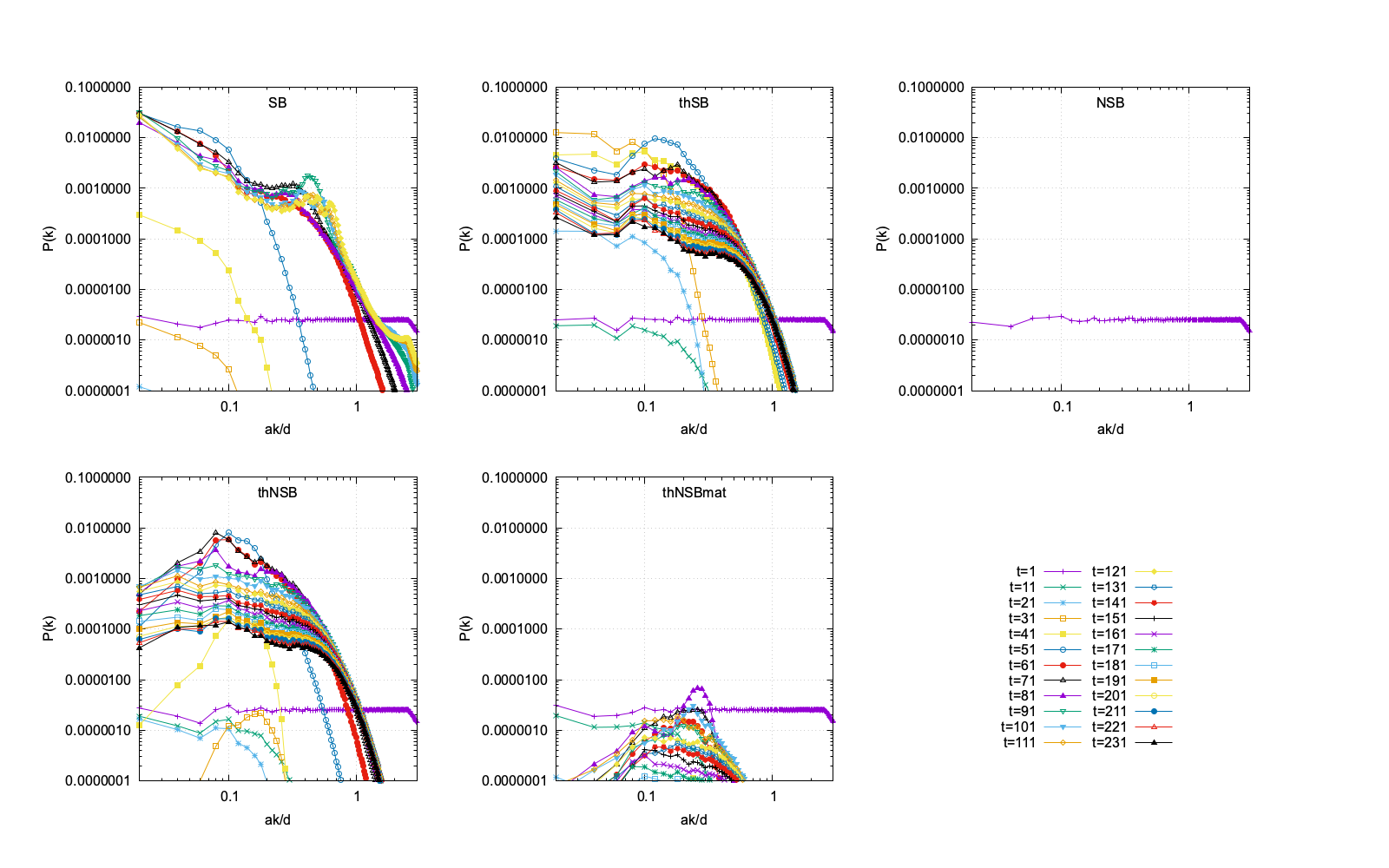}
\includegraphics[width=160mm]{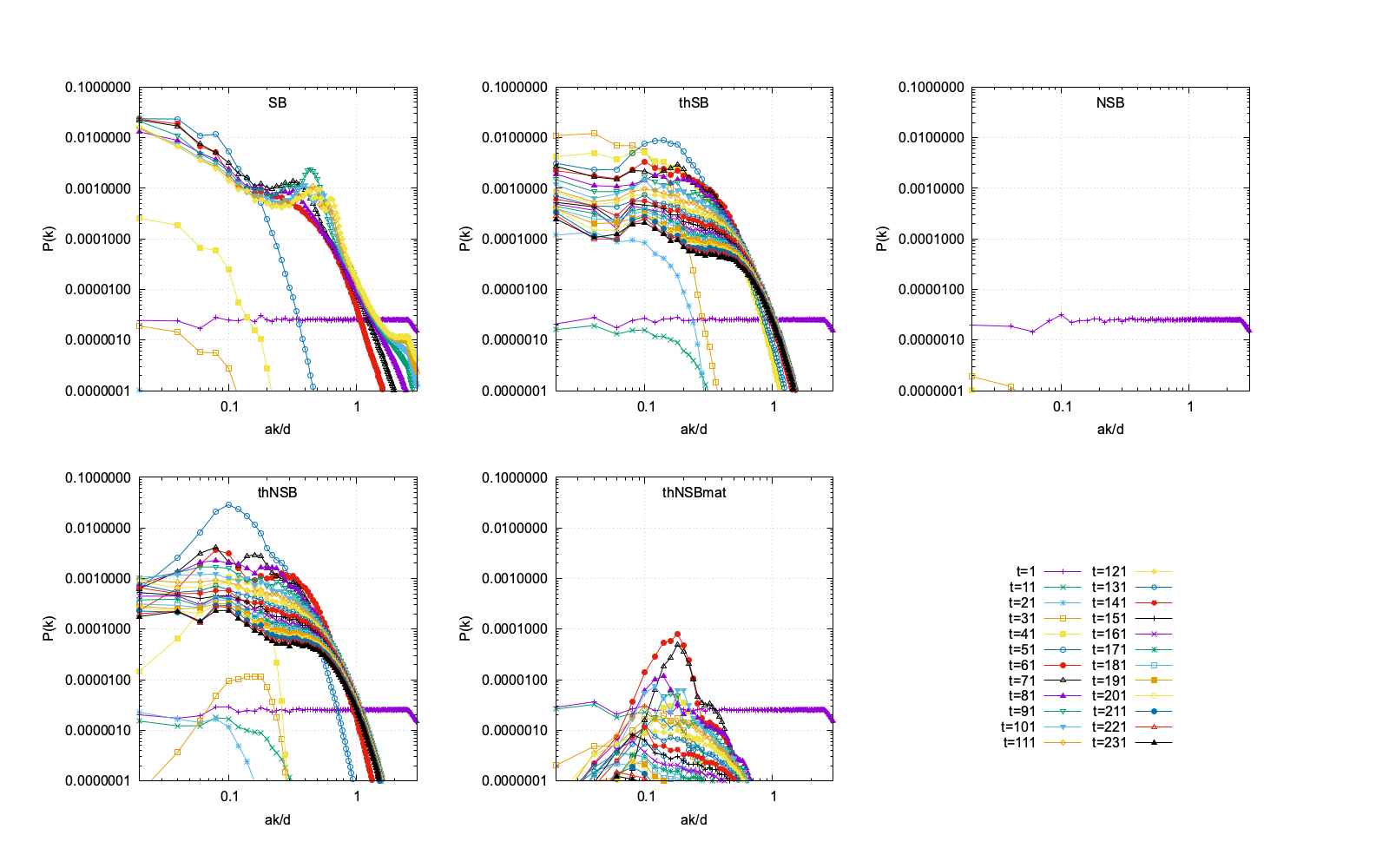}
    \end{center} 
    \caption{Reduced power spectrum of $\phi_1$ (top panel) and $\phi_2$ (bottom panel) in $U(1)$ models with scale-invariant fluctuations. $H_{\rm inf}^2 = 0.01d^2$.}
    \label{fig:U1scaleinvpslarge} 
\end{figure}

\bibliography{PQ5GW.bib}

\end{document}